# Shake-Table Tests of a Three-Storey Steel Structure with Resilient Slip-Friction Joints (RSFJ)

Nicholas Chan[1]; Ashkan Hashemi[2]; Soheil Assadi[3]; Pierre Quenneville[4]; Charles Clifton[5]; Gregory MacRae[6];

Rajesh Dhakal[7]; Liang-Jiu Jia[8]

**Abstract:** Shake-table tests were conducted as part of the Robust Building Systems (ROBUST) program to evaluate and demonstrate the resilience of various structural concepts under earthquake excitations. This paper presents the results of 3 different configurations incorporating the Resilient Slip-Friction Joint (RSFJ) in: (1) tension-only braces (TOB), (2) tension-compression braces (TCB), and (3) moment-resisting frame (MRF) joints along the longitudinal direction. Each building configuration was subjected to the El Centro ground motion at increasing intensities, reaching a peak ground acceleration (PGA) of about 0.5 g. The peak displacements sustained were well within design values, with the top floor deflecting by 0.66%, 0.98% and 1.26% of the total height for the TOB, TCB and MRF systems respectively. The corresponding inter-storey drifts peaked at 0.87%, 1.11% and 1.62%. In all cases, the structures essentially self-centered with residual drifts limited to 0.03%. The floor accelerations observed in unidirectional tests were 0.67 g (TOB), 0.66 g (TCB) and 0.62 g (MRF). The spikes may occur during (a) direction reversal at peak Mode 1 amplitude, (b) under the influence of higher Mode 2, and/or (c) stiffness transition at the upright/plumb position. In bidirectional tests, torsion effects contributed to the largest acceleration recorded at 0.95 g (TCB). Unbalanced resistance/deformation of symmetric friction connections in the transverse direction led to the structure twisting and loading the longitudinal RSFJs unevenly. However, the effects on peak displacements were not significant and similar responses as unidirectional tests indicate effective performance under simultaneous out-of-plane shaking. Overall, the results demonstrate the RSFJ's ability to damp seismically induced vibrations and restore structures to their original, undeformed shape after earthquakes. Avenues for future research include mitigating floor accelerations to further safeguard building contents and occupants.

## Background

Earthquakes in the 21[st] century (2000–2024) have already claimed over a trillion USD in damage and 800,000 lives globally (Kaneko 2024; TBMM 2023; UNDRR 2019; Uroš et al. 2024). The human costs of earthquakes are projected to grow in the foreseeable future because of increasing urbanization in earthquake-prone areas, vulnerability of the current building stock and inter-dependency of economies that lead to a ripple effect. In the US, FEMA P-366 estimates that the annualized earthquake loss to the building stock is US$14.7 billion per year (FEMA 2023). This is dominated by California which accounts for almost US$10 billion per year due to the seismic hazard and structural exposure. To highlight the seismic vulnerability, Ramirez et al. (2012) found that modern reinforced-concrete buildings in California are expected to sustain damage between 16% to 50% of the replacement cost – averaging at 32% – in an earthquake they were designed for. However, FEMA P-58-1 notes that the true costs are higher because most building owners will choose to demolish the building once the repair cost exceeds 40% of the replacement cost, with all other factors considered (FEMA 2018).

---
[1] Research Asst., Dept. of Civil and Environmental Engineering, Univ. of Auckland, Auckland, New Zealand. E-mail: jcha367@aucklanduni.ac.nz.
[2] Senior Lecturer, Dept. of Civil and Environmental Engineering, Univ. of Auckland, Auckland, New Zealand. E-mail: ahas439@aucklanduni.ac.nz.
[3] Ph.D. Candidate, Dept. of Civil and Environmental Engineering, Univ. of Auckland, Auckland, New Zealand. E-mail: skha858@aucklanduni.ac.nz.
[4] Professor, Dept. of Civil and Environmental Engineering, Univ. of Auckland, Auckland, New Zealand. E-mail: p.quenneville@auckland.ac.nz.
[5] Professor, Dept. of Civil and Environmental Engineering, Univ. of Auckland, Auckland, New Zealand. E-mail: c.clifton@auckland.ac.nz.
[6] Professor, Dept. of Civil and Natural Resources Engineering, Univ. of Canterbury, Christchurch, New Zealand. E-mail: gregory.macrae@canterbury.ac.nz.
[7] Professor, Dept. of Civil and Natural Resources Engineering, Univ. of Canterbury, Christchurch, New Zealand. E-mail: rajesh.dhakal@canterbury.ac.nz.
[8] Professor, Dept. of Disaster Mitigation for Structures, Tongji Univ., Shanghai, China. E-mail: lj_jia@tongji.edu.cn

An example of the scale of the damage and disruption was observed in New Zealand where 70% of buildings in the Christchurch city-center were demolished in the aftermath of the 2011 earthquake (Filippova et al. 2021). Even though very few buildings collapsed, many reinforced-concrete structures (especially RC frames) sustained distributed damage and residual deformation that were difficult to repair. Hence, there has been a shift towards other types of steel systems that are perceived as low damage and easier to repair (Bruneau and MacRae 2017).

Low damage concepts have progressed rapidly in the past few decades, offering more accessible solutions for low-to-medium rise buildings. Replaceable fuses or dissipators, like buckling-restrained braces, can reduce the damage to the gravity load-bearing system by confining most of the damage to the designated yielding components (Hussain and Kim 2023). An alternative to yielding dissipators are friction dampers. The friction can provide consistent and repeatable dissipation with minimal degradation after several earthquakes (Christopoulos et al. 2008; Tsai et al. 2008). While adequate dissipation helps to damp the vibrations during earthquakes, an increasingly recognized aspect of resilient buildings is the need to minimize residual deformations after earthquakes and the associated disruption to society (Filippova et al. 2021). Thus, research on self-centering systems has been increasing exponentially since 2014. Initial studies explored the use of post-tensioned rods to help buildings return to an upright position (Priestley et al. 1999; Ricles et al. 2002). Recently, researchers have been exploring self-centering dampers as alternatives to post-tensioned rods.

**Resilient Slip-Friction Joint (RSFJ)**

The Resilient Slip-Friction Joint (RSFJ) is one such damper designed to control seismic behavior at the joint level (Fig. 1). Conceived in 2015 as a low-damage and low-maintenance solution, it has already been adopted (or in the process of being implemented) in several earthquake prone regions like New Zealand, Canada, USA and Japan (Tectonus 2024). The joint is known for its ability to provide damping and self-centering passively in an integrated unit. This is achieved by using Belleville springs to provide restoring forces at the damper level, thus offering a more compact alternative to post-tensioned solutions.

The RSFJ has been investigated in rocking timber and concrete walls, and steel and timber braces (Fig. 2) (Bagheri et al. 2020; Hashemi et al. 2018; Shabankareh 2020; Yousef-beik et al. 2024). However, the tests performed on RSFJ systems to date are limited to quasi-static or dynamic tests on isolated parts of the lateral load-resisting systems (system level). This paper presents the results of the first shake table test of the RSFJ, which included both the complete vertical and lateral structural systems. One of the goals of the tests is to verify the self-centering capability of a full-scale structure and demonstrate its resilience against earthquake induced shaking.

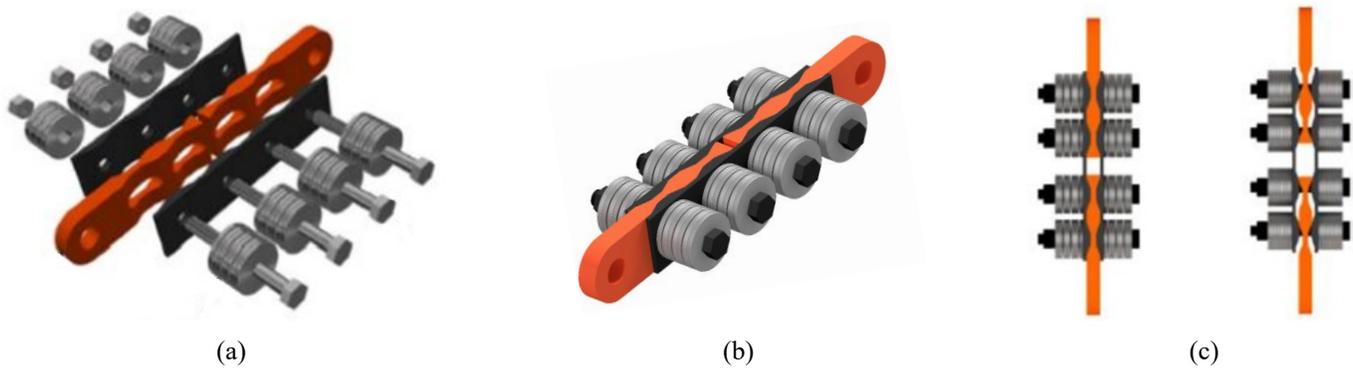

(a)                          (b)                          (c)

**Fig. 1.** Illustrating the RSFJ. (a) Exploded view. (b) Assembled and pre-stressed. (c) Loaded in tension.

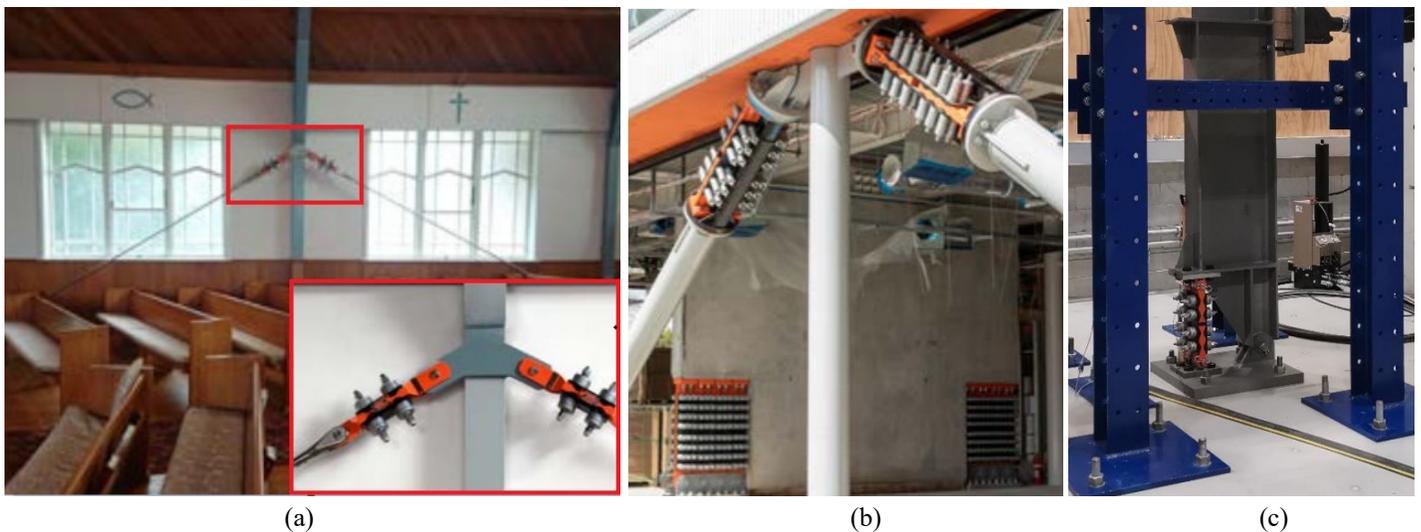

(a)                                  (b)                                  (c)

**Fig. 2.** RSFJ in (a) tension-only braces (TOB), (b) tension-compression braces (TCB) and (c) moment-resisting frame (MRF) joint.

## Test Structure

### Overview and Design

A three-storey steel structure was designed and constructed as part of the Robust Building Systems (ROBUST) shake-table test program which aims to demonstrate the resilience of several structural and non-structural concepts against earthquakes (Assadi et al. 2024; MacRae et al. 2020). The steel structure measures 7.25 m longitudinally by 4.75 m in the transverse direction and reaches 9 m tall at the top floor (Fig. 3). The structure was designed according to the New Zealand seismic code NZS 1170.5 (Standards New Zealand 2016) with the following design parameters:

- Located in the seismically active central business district of Wellington, New Zealand, at 5 km from the nearest fault,

- Building is designed for a 500-year return period earthquake, with 10% probability of exceedance in 50-year lifespan,

- Shallow soil conditions, i.e., shear wave velocity between 150–300 m/s and low-amplitude period under 0.6 sec.

                                                                                                  

The central column (200 x 200 x 10 SHS) was designed to take gravity loads only while the perimeter columns (250UC89.5) were intended to resist both gravity and earthquake loading. They were designed based on the larger of (a) earthquake actions assuming a structural ductility factor of 1.25, and (b) the overstrength capacity being reached in all the RSFJs. All columns have flexural continuity over their entire height. The beams (310UB40.4) were designed using the equivalent static method in NZS 1170.5 assuming a structural ductility of 3, in line with the strong-column weak-beam approach. The beams support a composite floor (Steel&Tube 2016) with trapezoidal steel deck and concrete topping. The floor is 150 mm at its thickest and has an effective thickness of 121 mm. Additional mass blocks were placed on each floor to simulate the imposed loads. Table 1 shows the seismic weights.

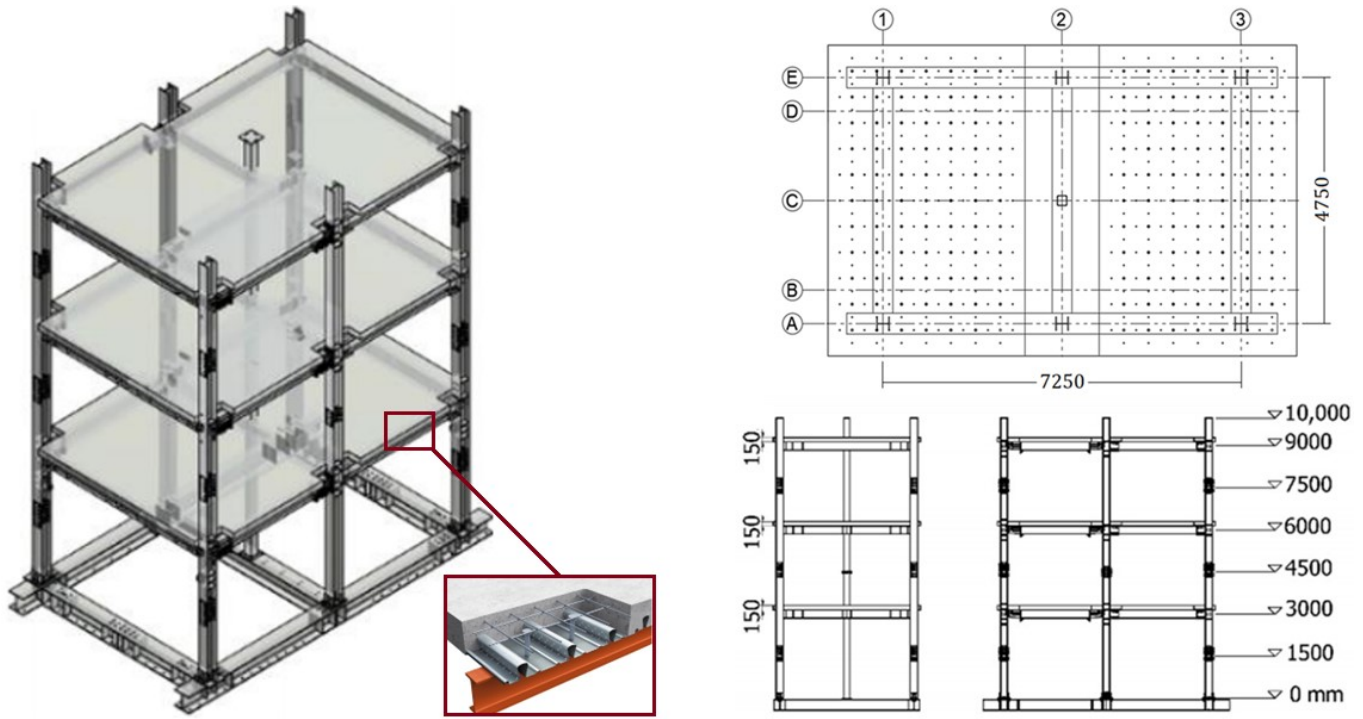

**Fig. 3.** Schematic of the test structure.

**Table 1.** Seismic weights.

| Storey | Height (m) | Seismic weight (kN) |
| --- | --- | --- |
| 3 | 9 | 276 |
| 2 | 6 | 251 |
| 1 | 3 | 251 |

The test structure was designed to accommodate testing of 8 different lateral load-resisting systems. This includes 3 types of moment-resisting frames, 4 types of concentrically braced frames and 1 rocking braced frame. The RSFJ concepts comprise 3 out of the 8 different systems (Fig. 4). These are (1) tension-only braces (TOB), (2) tension-compression braces (TCB), and (3) moment-resisting frames (MRF). The RSFJ concepts are used in one bay only between grid lines 1 and 2 to prevent detailing conflicts with the other concepts used in the other bay between grid lines 2 and 3. In the braced frames, the braces were used in the bottom two storeys only because the columns were strong enough to limit the inter-storey drift in the top storey.

These systems were designed to reach a peak roof displacement of 187 mm or 2.1% of the total height during the ultimate limit

                                                                                                                4

state (ULS) intensity of shaking, i.e., return period of 500 years. However, in practice the target displacement is often below 1.5% to limit the damage to drift-sensitive non-structural elements (Dhakal et al. 2016). The structure has additional capacity to reach a roof displacement of 337 mm or 3.7% of the total height during the maximum considered earthquake (MCE) with a return period of 2500 years. All the RSFJs were designed with sufficient deformation capacity to sustain a drift of 4.5%. To simplify the design, only one type of RSFJ was fabricated per system. Thus, the only difference between RSFJs in each system is the amount of pre-stressing for different storeys. Table 2 summarizes the design parameters and more information on the design is available from Bagheri (2022).

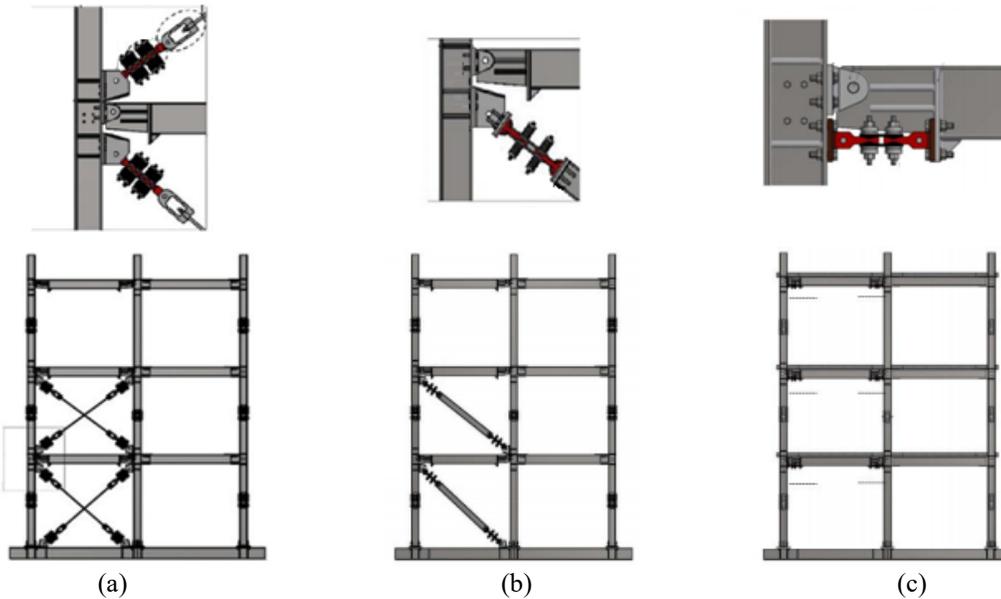

(a)                  (b)                  (c)

**Fig. 4.** Illustrating the three RSFJ configurations tested on the shake table. (a) TOB. (b) TCB. (c) MRF.

**Table 2.** Design parameters of the RSFJ lateral load-resisting systems in the longitudinal direction.

| Design Parameter | TOB | TCB | MRF |
|---|---|---|---|
| Mode 1 period, $T_1$ | 0.46 s | 0.48 s | 1.14 s |
| Mode 2 period, $T_2$ | 0.17 s | 0.17 s | 0.23 s |
| Mode 3 period, $T_3$ | 0.13 s | 0.15 s | 0.13 s |
| Base shear at slip, $V_{b, slip}$ | 158 kN | 156 kN | 83 kN |
| Base shear at ult., $V_{b, ult}$ | 272 kN | 264 kN | 115 kN |

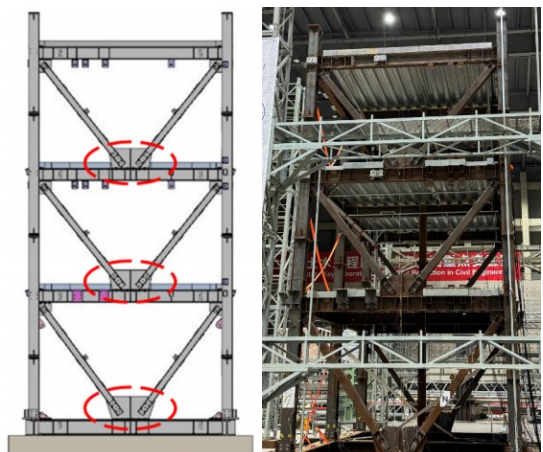

**Fig. 5.** Concentric V-braces with symmetric friction connections (SFC) in the transverse direction.

The transverse direction utilizes concentric V-braces as shown in Fig. 5. The brace bodies are made from a pair of rectangular hollow sections (each measuring 160 x 80 x 6 mm) that were joined back-to-back by welding steel plates onto the top and bottom flanges. The connections are fixed at the top end of the braces. At the bottom end of the braces, symmetric friction connections (SFC) with Belleville springs are used to resist transverse loading. The slip forces in the SFC are approximately 100 kN (Yan et al. 2020).

***General Connections***

This section describes other connections that are part of the seismic-resisting system: namely, the beam-to-column connections and the column-to-foundation connections. They were designed based on the Structural Steel Connections Guide by Steel Construction New Zealand with some modifications. First, the connections between gravity beams and columns were different in each of the two bays (Fig. 6) to accommodate the different systems tested. Thus, only one bay resisted lateral loads at any point in time.

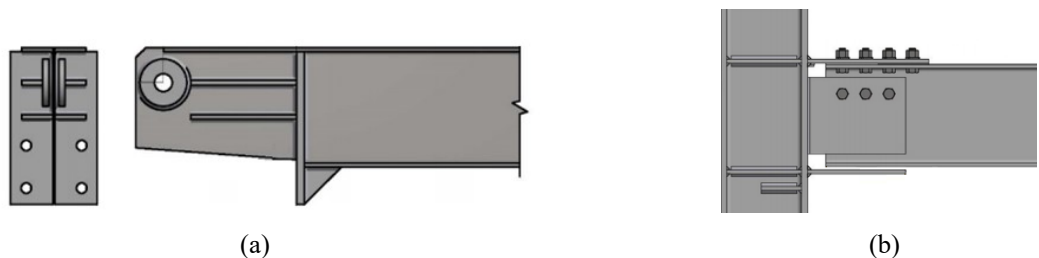

(a)                                                            (b)

**Fig. 6.** Beam-to-column detail in (a) left bay for the RSFJ concepts, (b) right bay for other concepts.

In the bay used by the RSFJ concepts (grid lines 1-2), the beams were connected to the columns by simple pins to transfer the gravity loads to the columns. Additional moments are also resisted when the RSFJ is used in the MRF configuration. When the RSFJ bay is active, there must be negligible resistance from the other bay which is intended for other concepts in the ROBUST program like the sliding hinge joint. Hence, the beams in the other bay were bolted on the top flange only and disconnected at the bottom flange. This allows freedom of rotation that is compatible with the swaying structure, while supporting the gravity loads still.

Fig. 7 shows the typical connection between column base and foundation beam, which uses an asymmetric friction connection (Borzouie et al. 2016). The original design intent of the friction side-plates was to allow for different levels of clamping forces, and therefore the option of either nominally pinned or nominally fixed column bases. During a strong earthquake (at ULS level) sliding would occur in the friction connection and allow rotational freedom. At lower intensities of shaking (below SLS), they were intended to remain fastened rigidly like fixed connections. However, it was realized at a later stage of the design process that the friction plates were not necessary to restrict the rotation at low levels of shaking. Thus, the friction plates were not used for the RSFJ concepts.

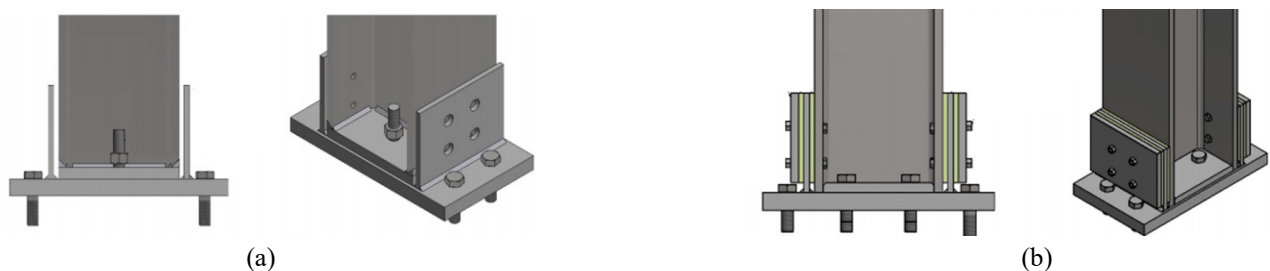

(a)                                                            (b)

**Fig. 7.** Column base (a) nominally pinned without friction plates, and (b) nominally fixed with friction plates.

                                                                                              

**Three RSFJ Systems**

*System 1: Tension-Only Braces (TOB)*

Tension-only braces are often used in low-rise industrial and commercial steel buildings due to their affordability and simplicity in design. They can be very efficient because no buckling resistance or restraints are needed in the brace. In ordinary braces, a degree of slackness grows over successive cycles of yielding in tension (e.g. after a large pulse or inelastic excursion). The pinched hysteresis loops impair the stiffness and dissipation capabilities of the system. Because of this, various codes have restrictions on the use of tension-only braces such as maximum slenderness limits (Mehrabi et al. 2019). Recently, ratcheting devices have been proposed for tension-only braces to restore the dissipation while providing self-centering at the same time (Chan et al. 2023). However, even if most of the slack is eliminated, a small amount of backlash remaining in the ratchet impairs the initial stiffness of the system.

The RSFJ has been investigated for tension-only braces, demonstrating self-centering successfully and repeatedly without deterioration (Bagheri et al. 2020). This test incorporates some of the findings from that study (Fig. 8). A main consideration is the need to minimize the sagging of the braces under the self-weight of both RSFJ and the rod. This is important because the sagging results in a slackness in the system. The slackness leads to an undesirable delay in mobilizing the full lateral load resistance and it reduces the initial stiffness of the system. Thus, the braces are tightened to apply some pre-load for tautness. The pre-loading subjects the frame to self-balancing internal stresses from the cross-braces. However, the pre-loading applied on the joints reduces the lateral resistance available before the joints start to slip. Hence, the joints in this study are designed to a higher slip load to account for the additional preload. The braces are pinned on both ends and use 25 mm rods.

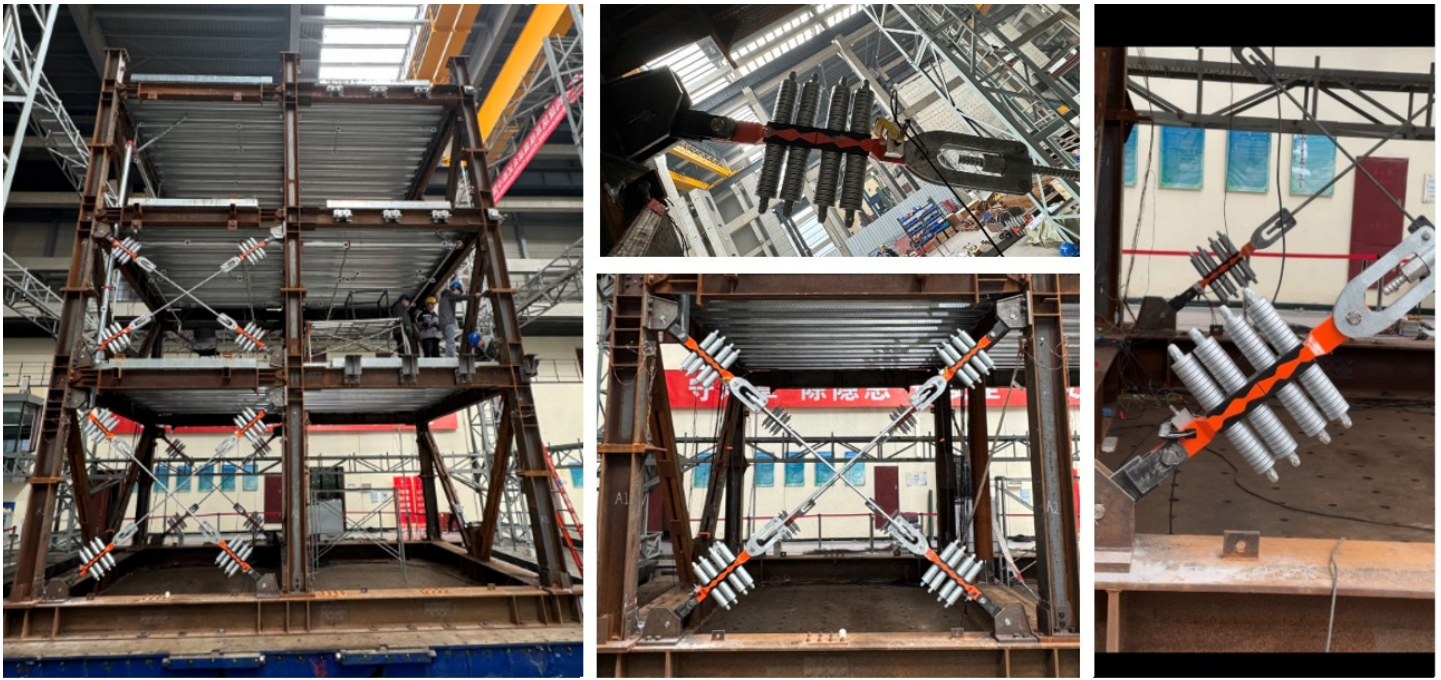

**Fig. 8.** Tension-only brace (TOB) system.

***System 2: Tension-Compression Braces (TCB)***

Unlike TO cross-braces, only one TC diagonal brace is needed per bay as the brace undergoes tension or compression depending on the direction of load. Self-centering characteristics can be introduced into individual braces by incorporating springs or restoring force mechanisms in parallel with a buckling-restrained or buckling-resistant TC brace (e.g., BRB with a yielding core, or a stocky brace member with ordinary friction sliders). The types of springs include post-tensioned strands (Christopoulos et al. 2008; Erochko et al. 2015), friction springs (Filiatrault et al. 2000) and disc springs (Zhang et al. 2022). In these cases, a double-acting, telescoping configuration is used to trigger the restoring forces by loading the spring mechanism elastically regardless of the direction of load on the brace. In the examined T/C brace, a telescoping mechanism is not needed to load the disc springs in the RSFJs – this is achieved by the symmetric groove profile. Smaller telescoping tubes are used to control buckling in the joint instead.

As the RSFJ evolves from initial prototypes into more slender modular units, buckling becomes a primary aspect of the design that has been observed in past studies and needs to be controlled (Yousef-beik et al. 2024). For the TCB, it includes both joint buckling (i.e., joint rotates internally) and brace buckling (elastically and inelastically). At the joint level, the buckling is prevented by using anti-buckling tubes (ABT) which are hollow sections that telescope (to allow axial deformation) while resisting rotation. The design method to control brace buckling uses a simplified method governed by (1) elastic Euler buckling load reduced by the presence of rotational springs, and (2) the inelastic yielding within the brace body (200UC59.5). Both the initial imperfections and load-dependent eccentricity are considered when calculating the buckling capacity. The resulting ABT in Fig. 9 has inner and outer diameters of 30 mm and 50 mm for the male portion, and 52 mm and 65 mm for the female portion.

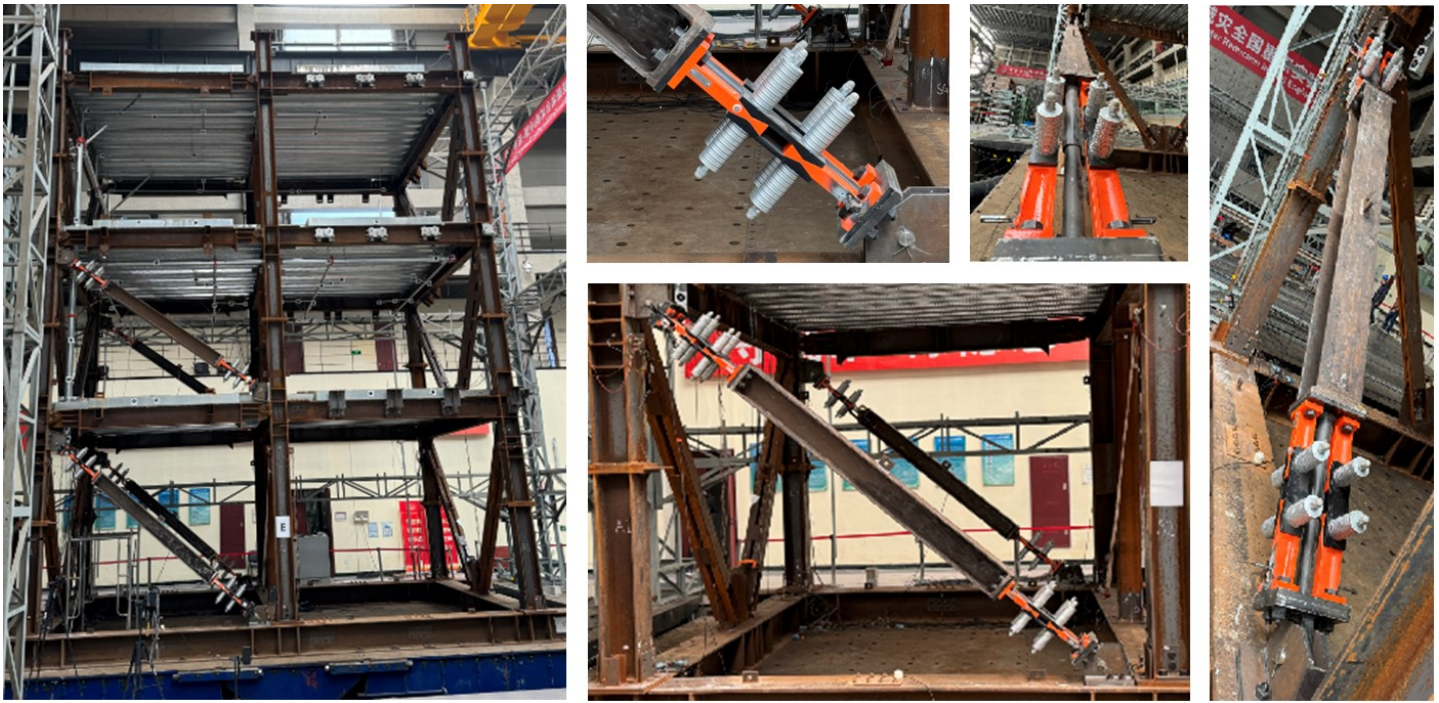

**Fig. 9.** Tension-compression brace (TCB) system.

### System 3: Moment-Resisting Frame (MRF)

Moment-resisting frames offer greater architectural freedom with unobstructed views. As the building sways, critical regions at the beam-column joints resist the relative rotation between beam and column. Brittle fractures at the beam-to-column welds have been observed during the Northridge earthquake in 1994. Since then, ductile and low-damage solutions have been developed in line with the strong-column weak-beam approach. Some of the plastic or dissipative mechanism includes reduced beam sections and slotted plates for dissipation by friction (Shabankareh 2020). Additional use of post-tensioned tendons and/or friction springs have been explored as a way to re-center both the joint and frame after earthquakes (Ricles et al. 2002). Researchers have investigated the use of dissipators at the bottom flange only, such that the center of rotation occurs where the top flange meets the column face (MacRae et al. 2010). This minimizes gap-opening at the top flange (frame expansion) and the associated damage to the overlying floor slab.

The RSFJ is similarly applied at the bottom flange (Fig. 10). A moment resisting connection is created with a lever arm of 220 mm, offset vertically from a large-diameter pin at the top. The pin which serves as the center of rotation is also intended to support gravity loads. Thus, the beam-end regions had to be reinforced via stiffeners and thickened webs to accommodate the shearing, bending and bearing forces through the pin joint. Similarly, column webs were stiffened in the vicinity of the force couple to mobilize the entire depth of the section. The RSFJs have different number of discs per side to allow room for stiffeners at the beam-ends.

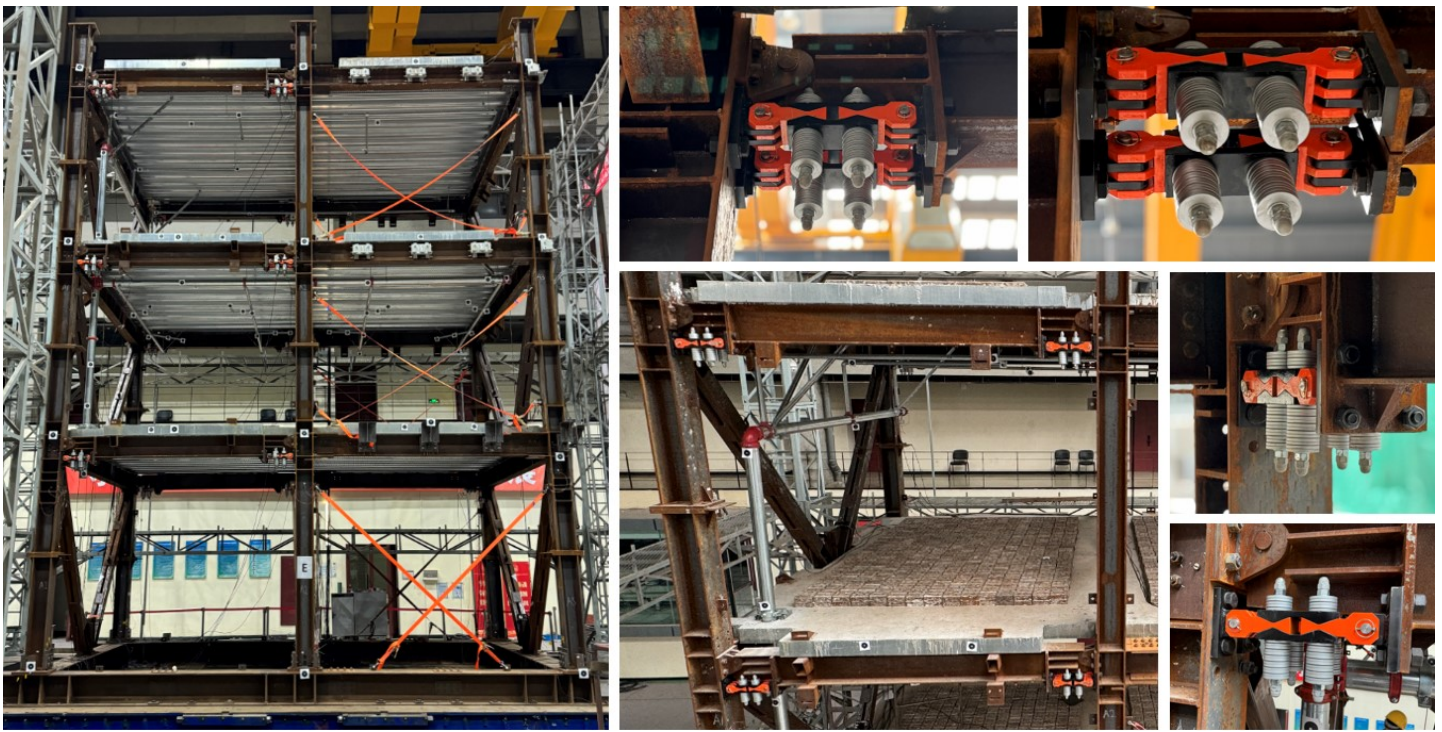

**Fig. 10.** Moment-resisting frame (MRF) system.

*RSFJ Cyclic Behavior*

All the RSFJs used in the tests were subjected to preliminary cyclic testing. Given that the joints were manually pre-stressed, the forces were allowed to vary from joint-to-joint up to a maximum of 10%. Representative test results for the RSFJ hysteresis are shown in Figs. 11, 12 and 13 for the TOB, TCB and MRF systems respectively. The average speed achieved with a manually operated hydraulic pump was approximately 2 mm/s. However, previous tests have demonstrated rate independence up to 100 mm/s (Bagheri et al. 2020) and more recently 160 mm/s as per ASCE 7-16 requirements (Tectonus 2024).

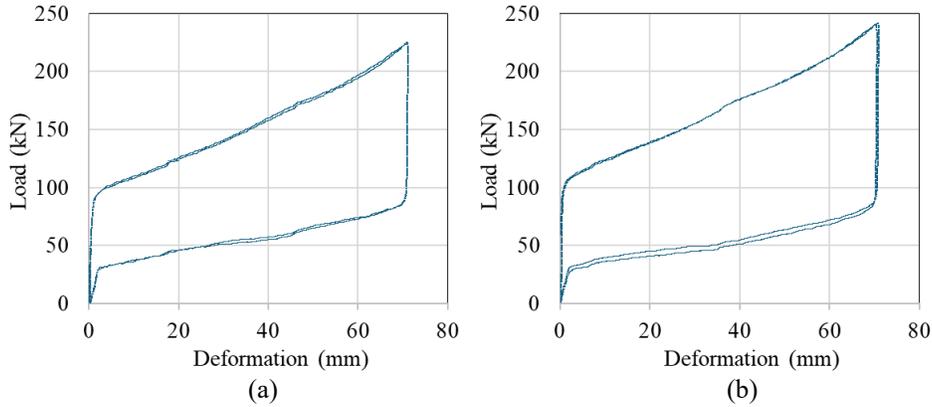

**Fig. 11.** Typical test result for the RSFJs used in the TOB system. (a) First and (b) second storeys.

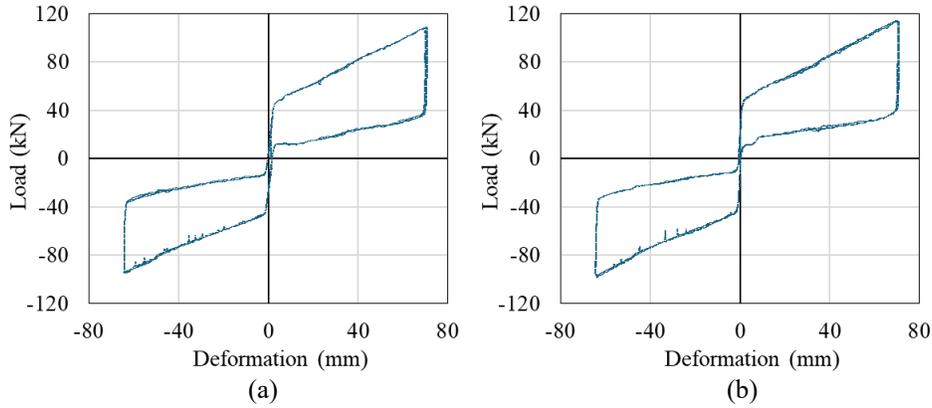

**Fig. 12.** Typical test result for the RSFJs used in the TCB system. (a) First and (b) second storeys.

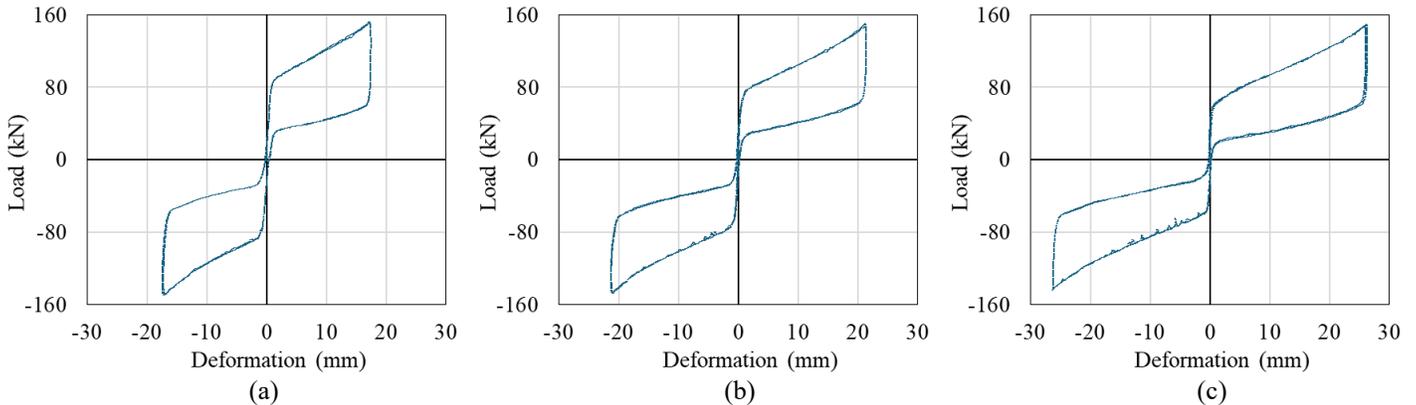

**Fig. 13.** Typical test result for the RSFJs used in the MRF system. (a) First, (b) second and (c) third storeys.

**Instrumentation**

The global response of the structure was monitored with a combination of potentiometers and accelerometers. The accelerometers used are Setra 141 by Setra Systems with a measurement range up to +/- 8g. A range of displacement sensors with different stroke lengths were used from TM Automation Instruments. They include linear potentiometers NS-WY02 (50 mm stroke) and NS-WY03 (100 mm stroke), and string potentiometers NS-WY06 (from 200 mm to 1000 mm stroke) and NS-WY07 (1500 mm stroke).

Fig. 14 shows the layout of the instruments on a typical floor. String potentiometers, attached to a reference structure, were used to measure the absolute displacements at two opposite corners (columns A1 and E3) for all levels. Accelerometers were placed on all 3 storeys at the two corner columns (A3 and E1) as well as the center of each floor (C2). On the ground level, accelerometers were also placed on the shake table and the foundation beam. Linear potentiometers were also used to check for any unintended movement between the foundation beam and the shake table.

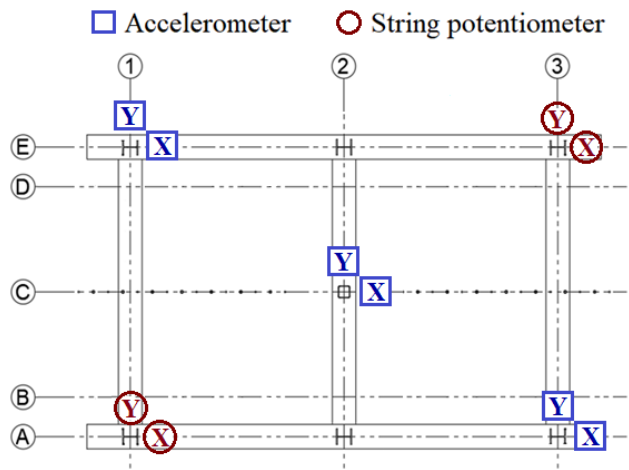

**Fig. 14.** Instrumentation layout on a typical floor for measuring global structural responses.

The deformations of the RSFJs were measured with string potentiometers (Fig. 15). In the transverse direction, V-braces were instrumented with linear potentiometers to capture any relative slip in the symmetric friction connection (i.e., slip between the brace and the gusset). An additional linear potentiometer was used to check for any unintended sliding at the top (fixed) end of the first storey braces. The sampling rate was 256 Hz which is approximately one data point every 0.0039 second.

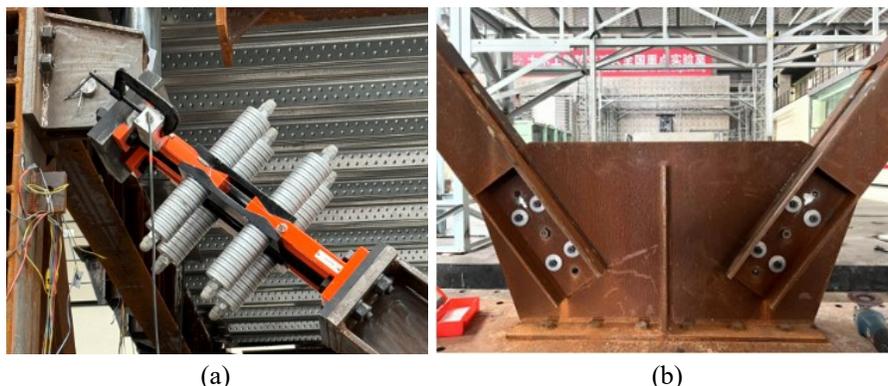

(a)                                          (b)

**Fig. 15.** Instrumenting the (a) RSFJ (longitudinal) and (b) SFC (transverse) with string potentiometers.

**Ground Motions and Test Sequence**

Several structural and non-structural concepts were investigated during the ROBUST program. Because of this, the total number of runs can be considerably high with a full suite of ground motions and quickly become unfeasible in terms of time and budget constraints. Hence, all systems were subjected to one ground motion scaled to multiple intensities. After a subjective quantitative assessment, the North-South component of the 1940 El Centro ground motion was selected as shown in Fig. 16 and Table 3.

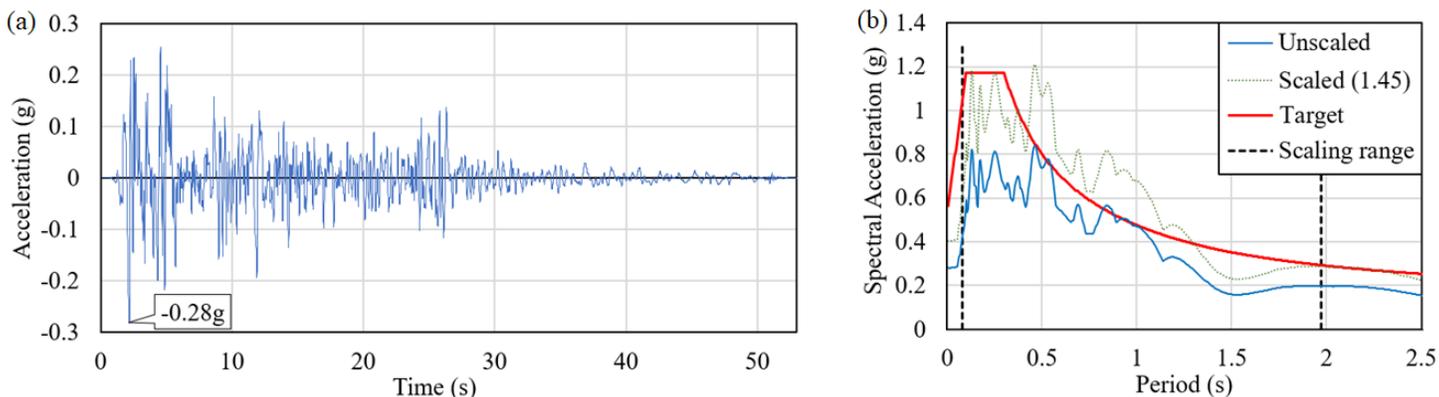

**Fig. 16.** North-South component of the 1940 El Centro (a) accelerogram and its (b) acceleration spectrum.

**Table 3.** Ground motion metadata as per the PEER NGA-West2 strong motion database.

| Event | Station | Year | $M_W$ | $R_{JB}$ (km) | $V_{s,30}$ (m/s) | Fault | File name |
|-------|---------|------|-------|---------------|------------------|-------|-----------|
| Imperial Valley | El Centro Array #9 | 1940 | 6.95 | 6.09 | 213 | Strike-slip | RSN6_IMPVALL.I_I-ELC180.AT2 |

The main reasons for selecting the 1940 El Centro ground motion include: (1) reasonable fit with design spectrum and site conditions such as magnitude, distance to fault, site subsoil, etc., (2) shake table can accommodate the displacement, velocity and acceleration demands imposed by the scaled records, (3) lab technical staff have prior experience working with this record. On a scale of 0 to 1, the pulse indicator for this ground motion is 0.09 which indicates it is unlikely to have a pulse (Baker 2007). The pulse indicator was calculated via a continuous wavelet transform that uses the Daubechies wavelet of the 4th order.

Table 4 shows the sequence of tests performed and the set of scale factors corresponding to the different intensities. The period range used to scale the ground motions starts from 0.4 times the shortest translational period $T_{1, min}$ to $\sqrt{\mu}$ times the longest translational period $T_{1, max}$ among all the systems tested. This translates to a period range between 0.18 sec and 1.97 sec which accounts for higher mode effects and period-lengthening effects resulting from the nonlinear behavior. Eq. 1 shows the formula used to calculate the scale factor $SF$ that provides the best fit with the target spectra. The scale factor minimizes the sum of squared errors, in terms of spectral accelerations $S_a$, between the scaled spectrum $SF * S_a^{unscaled}$ and the target spectrum $S_a^{target}$ over the nominated range of periods $T_i = 1 \dots N$ (Zhang et al. 2020). Thus, the design intensity requires scaling by 1.45 to reach the ultimate limit state (ULS) level of shaking (e.g., Target PGA = 1.45 * 0.28 g = 0.4 g).

**Table 4.** Sequence of tests performed.

| Test # | Intensity (SLS=ULS/4) | Input ($x$) | Input ($y$) | Scale Factor | Target PGA (g) |
|---|---|---|---|---|---|
| 1 | WN | White noise | White noise | - | 0.05 |
| 2 | 0.8 SLS | El Centro 1940 | - | 0.29 | 0.08 |
| 3 | WN | White noise | White noise | - | 0.05 |
| 4 | 1.0 SLS | El Centro 1940 | - | 0.36 | 0.10 |
| 5 | WN | White noise | White noise | - | 0.05 |
| 6 | 2.0 SLS | El Centro 1940 | - | 0.73 | 0.20 |
| 7 | WN | White noise | White noise | - | 0.05 |
| 8 | 1.0 ULS | El Centro 1940 | - | 1.45 | 0.40 |
| 9 | WN | White noise | White noise | - | 0.05 |
| 10 | 1.2 ULS | El Centro 1940 | - | 1.74 | 0.49 |
| 11 | WN | White noise | White noise | - | 0.05 |
| 12 | 1.0 SLS | El Centro 1940 | El Centro 1940 | 0.36 | 0.10 |
| 13 | WN | White noise | White noise | - | 0.05 |
| 14 | 2.0 SLS | El Centro 1940 | El Centro 1940 | 0.73 | 0.20 |
| 15 | WN | White noise | White noise | - | 0.05 |
| 16 | 1.0 ULS | El Centro 1940 | El Centro 1940 | 1.45 | 0.40 |
| 17 | WN | White noise | White noise | - | 0.05 |
| 18 | 1.2 ULS | El Centro 1940 | El Centro 1940 | 1.74 | 0.49 |
| 19 | WN | White noise | White noise | - | 0.05 |

Note: Tests 12–19 performed for TCB and MRF systems only.

$$SF = \frac{\sum_{i=1}^{N}\left[S_a^{unscaled}(T_i) \cdot S_a^{target}(T_i)\right]}{\sum_{i=1}^{N}\left[S_a^{unscaled}(T_i)\right]^2} \qquad (1)$$

Some deviation was observed between the input motion required and the actual acceleration history recorded by the accelerometer at ground level. While the peak acceleration required was 0.40 g at the design (ULS) intensity, the peaks recorded during the tests were larger at 0.49 g, 0.41 g and 0.43 g for the TOB, TCB and MRF systems respectively. Furthermore, there were fluctuations in acceleration observed for all cases as Fig. 17 shows.

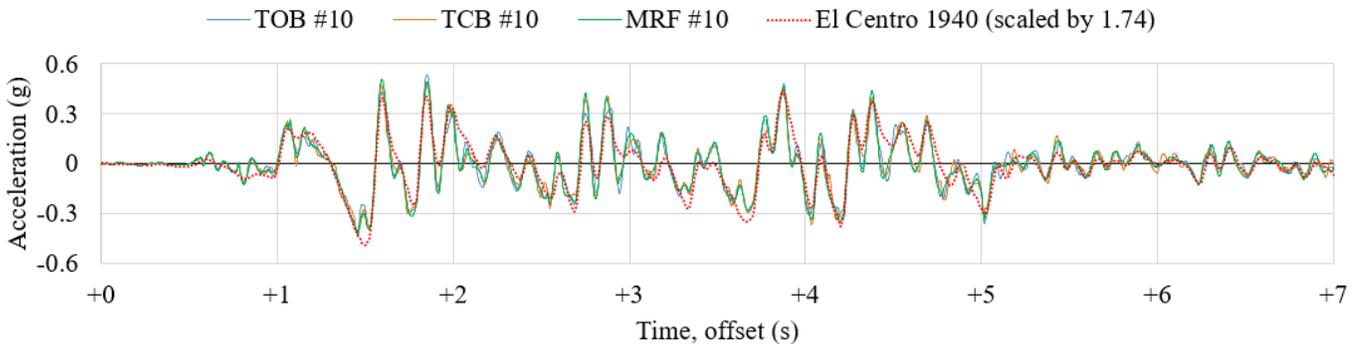

**Fig. 17.** Shake table acceleration history recorded during Test #10.

Given that the input ground motion deviates from the required El Centro motion, the acceleration spectra are recalculated and shown in Fig. 18 using the actual readings of the accelerometer placed on the shake table. These readings are obtained from the actual tests on all 3 systems, whose results are presented subsequently. Despite the differences in spectral shape from the original El Centro motion, the actual excitation provides a reasonable match with the target or design (ULS) spectrum up to a period of about 1.2 sec.

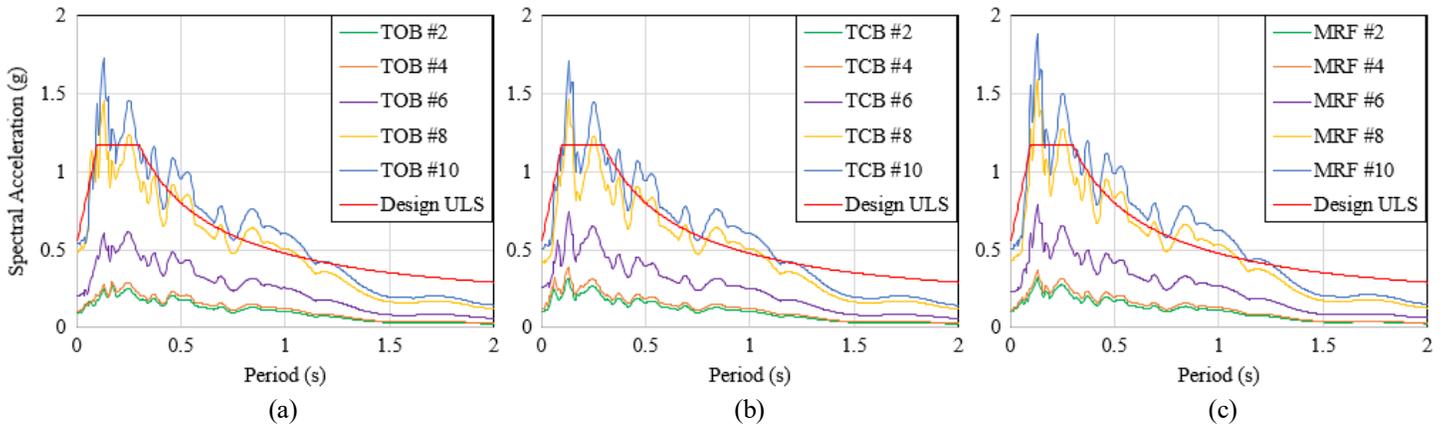

**Fig. 18.** Acceleration spectra of the excitation recorded on the shake table for (a) TOB (b) TCB and (c) MRF systems.

The white noise is depicted in Fig. 19(a) which shows the accelerometer readings recorded on the shake table during 100 seconds of excitation. The target PGA for the white noise was 0.05 g and root mean square amplitude was approximately 0.01 g. These low levels of excitation were intended to avoid any yielding in the structure. The signal appears to excite a range of frequencies as seen in a spectrogram in Fig. 19(b). The Fourier transform shows a relatively even range of amplitudes up to a frequency of 50 Hz. This upper limit is sufficient to capture any potential response to higher modes.

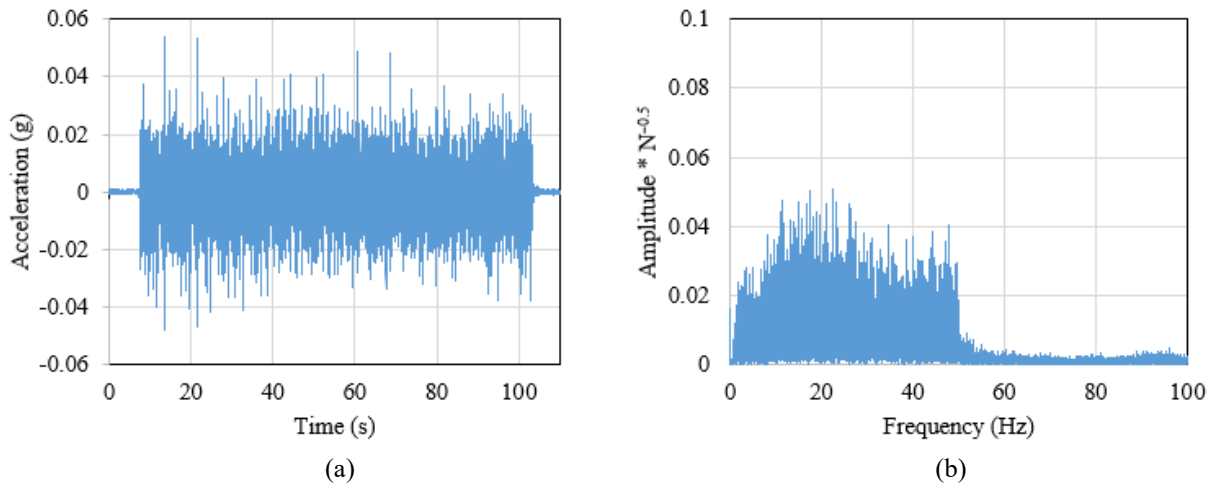

**Fig. 19.** (a) White noise excitation and (b) corresponding Fourier transform.

**Results and Discussion**

*Modal Identification*

The fundamental periods obtained from the tests were approximately 0.4 sec, 0.8 sec and 0.8 sec for the TOB, TCB and MRF concepts respectively. Fig. 20 shows the fundamental periods obtained from the acceleration responses. For tests with odd numbers (i.e., white noise excitation), the full response history was used to perform a frequency domain decomposition. For tests with even numbers (i.e., scaled El Centro), a Fourier transform was performed only for the free vibration portion during which the table is effectively stationary and the oscillations decay.

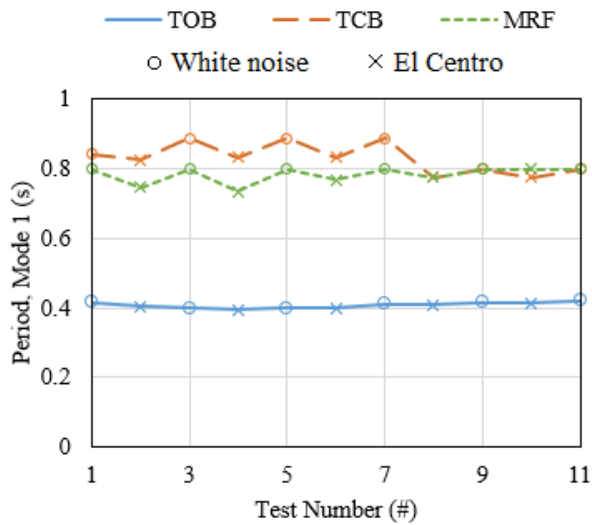

**Fig. 20.** Fundamental periods of the TOB, TCB and MRF systems.

The mode shapes and corresponding periods in Fig. 21 were identified from the acceleration responses to white noise excitation via frequency domain decomposition (Brincker et al. 2001). The shape vectors show 3 vibration modes in the longitudinal direction (X1, X2, X3), transverse direction (Y1, Y2, Y3) and twisting or torsional direction (T1, T2, T3). However, Mode T2 appeared to be relatively weak in comparison to the other modes on the power spectral density plot in Fig. 21(a). The mode shapes and periods are similar between the TCB and MRF systems, while the periods were smaller and more closely spaced for the TOB system.

Some differences are noted when comparing the periods obtained from the tests (Table 5) against the design values. For the TOB concept, tightening the cross-braces may have pre-stressed the steel frame to some extent, increasing the initial stiffness and therefore producing a lower fundamental period at 0.40 sec (test) against 0.46 sec (design). For the TCB concept, a small amount of slack was noted in the actual test set-up when installing the braces which are pinned on both ends and have a fixed length. Due to construction tolerances, it was not possible to match the length of the diagonal precisely (i.e., the distance between pin holes). Thus, the clearance and associated slack could have contributed to a higher-than-expected period of 0.80 sec (test) versus 0.48 sec (design). For the MRF concept, the period obtained was smaller than intended at 0.80 sec (test) against 1.14 sec (design).

**Table 5.** Modal periods obtained from frequency domain decomposition.

| System | Mode X1 | Mode X2 | Mode X3 | Mode Y1 | Mode Y2 | Mode Y3 | Mode T1 | Mode T2 | Mode T3 |
|--------|---------|---------|---------|---------|---------|---------|---------|---------|---------|
| TOB    | 0.410   | 0.154   | 0.059   | 0.243   | 0.062   | 0.038   | 0.143   | 0.041   | 0.026   |
| TCB    | 0.800   | 0.200   | 0.086   | 0.250   | 0.062   | 0.038   | 0.163   | 0.041   | 0.026   |
| MRF    | 0.800   | 0.191   | 0.084   | 0.235   | 0.062   | 0.038   | 0.160   | 0.043   | 0.026   |

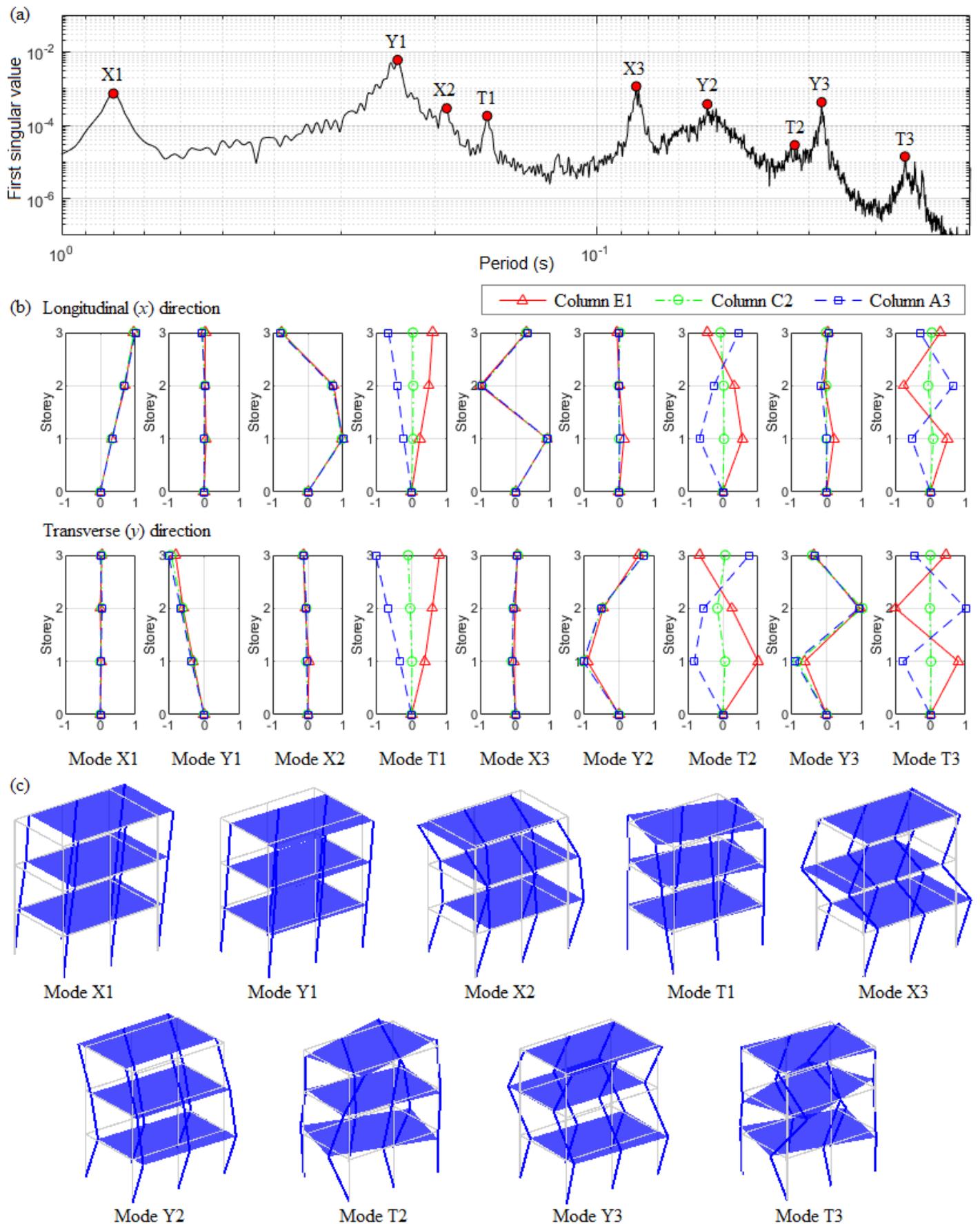

**Fig. 21.** Modal characteristics of the test structure. (a) Power spectral density. (b) Mode shape vectors. (c) 3D plots.

Fig. 22 shows 10 seconds of vibrations towards the end of Test #10 for all 3 systems. The amplitude of vibrations fit very well with the exponential decay function for the TOB system with a damping ratio of 1.15%. This is calculated from the logarithmic decrement equations in Eqs. 2 and 3 and Table 6 shows the parameters used to compute the envelopes. However, the exponential form results in a poorer fit for the other systems (TCB: 4.0% and MRF: 2.3%) as the envelope does not decay fast enough for the progressive cycles. Attempting to fit the envelope to the smaller trailing cycles would result in an underestimate of the initial larger cycles. The peaks of the larger cycles have a greater tendency to decrease more linearly (e.g., TCB system) indicating a possible source of frictional dissipation in the TCB system, and to a lesser extent in the MRF system. Hence, the true damping ratios attributed to viscous damping are likely to be in between 1.15% and 2.3%.

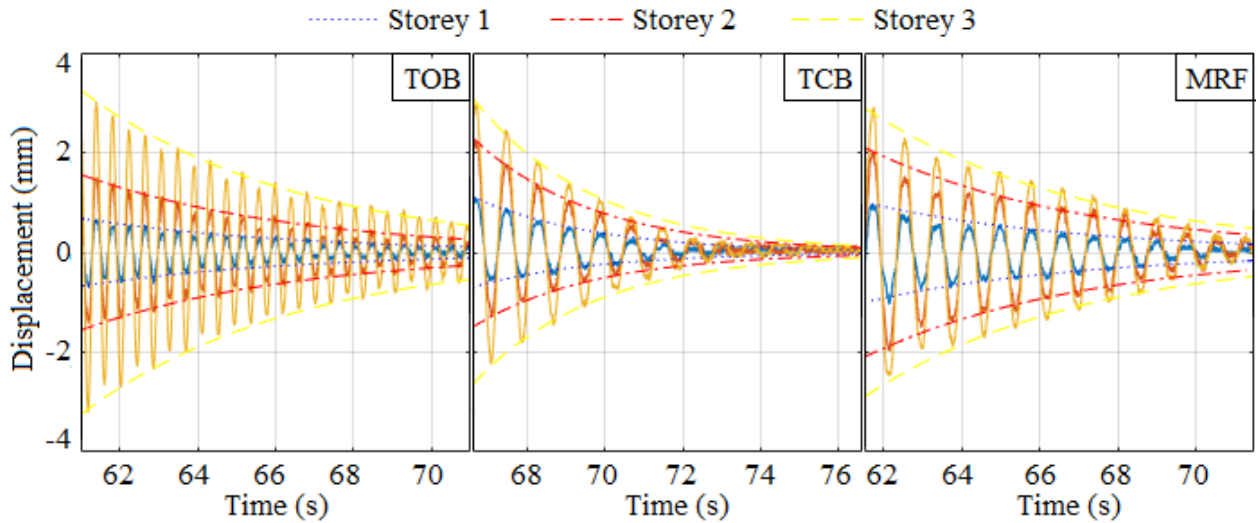

**Fig. 22.** Decay of vibrations over 10 seconds during Test #10 of the TOB, TCB and MRF systems.

$$\zeta = \frac{d}{\sqrt{4\pi^2 + d^2}} \tag{2}$$

$$d = \ln\left(\frac{x_n}{x_{n+1}}\right) = \ln\left(\frac{Ae^{-at}}{Ae^{-a(t+T)}}\right) = e^{aT} \tag{3}$$

**Table 6.** Parameters used to calculate the exponential decay envelope (i.e., logarithmic decrement).

| Sys. | Test # | $T$ (s) | $a$ | $d$ | $\zeta$ (%) | $A_1^+$ | $A_2^+$ | $A_3^+$ | $A_1^-$ | $A_2^-$ | $A_3^-$ |
|------|--------|---------|------|-------|-------------|---------|---------|---------|---------|---------|---------|
| TOB | 10 | 0.40 | 0.18 | 0.072 | 1.15 | 0.68 | 1.56 | 3.25 | −0.68 | −1.56 | −3.25 |
| TCB | 10 | 0.79 | 0.32 | 0.252 | 4.01 | 1.12 | 2.30 | 3.10 | −0.70 | −1.50 | −2.65 |
| MRF | 10 | 0.80 | 0.18 | 0.145 | 2.30 | 1.00 | 2.10 | 2.90 | −1.00 | −2.10 | −2.90 |

One observation is a minor asymmetry in the vibrations of the TCB system where the positive-x amplitudes are slightly larger compared to the negative-x amplitudes, with respect to the resting position. The asymmetry was not observed in prior tests of the TOB system, nor subsequent tests of the MRF system. Although the positive-x direction corresponds with the direction that subjects the braces to compression, the level of deformation is not very high to have activated the braces. Rather, it is believed that the slack due to construction tolerances in the TCB system may have been larger in the positive-x direction relative to the resting position.

*Overview of Global Response*

The peak responses to the scaled ground motions are depicted in Fig. 23 in terms of the (a) peak absolute accelerations, (b) peak displacements relative to the table, and (c) peak inter-storey drifts. The residual inter-storey drifts are included in Table 7, which summarizes the global response. More detailed values from individual instruments are available in the supplemental materials.

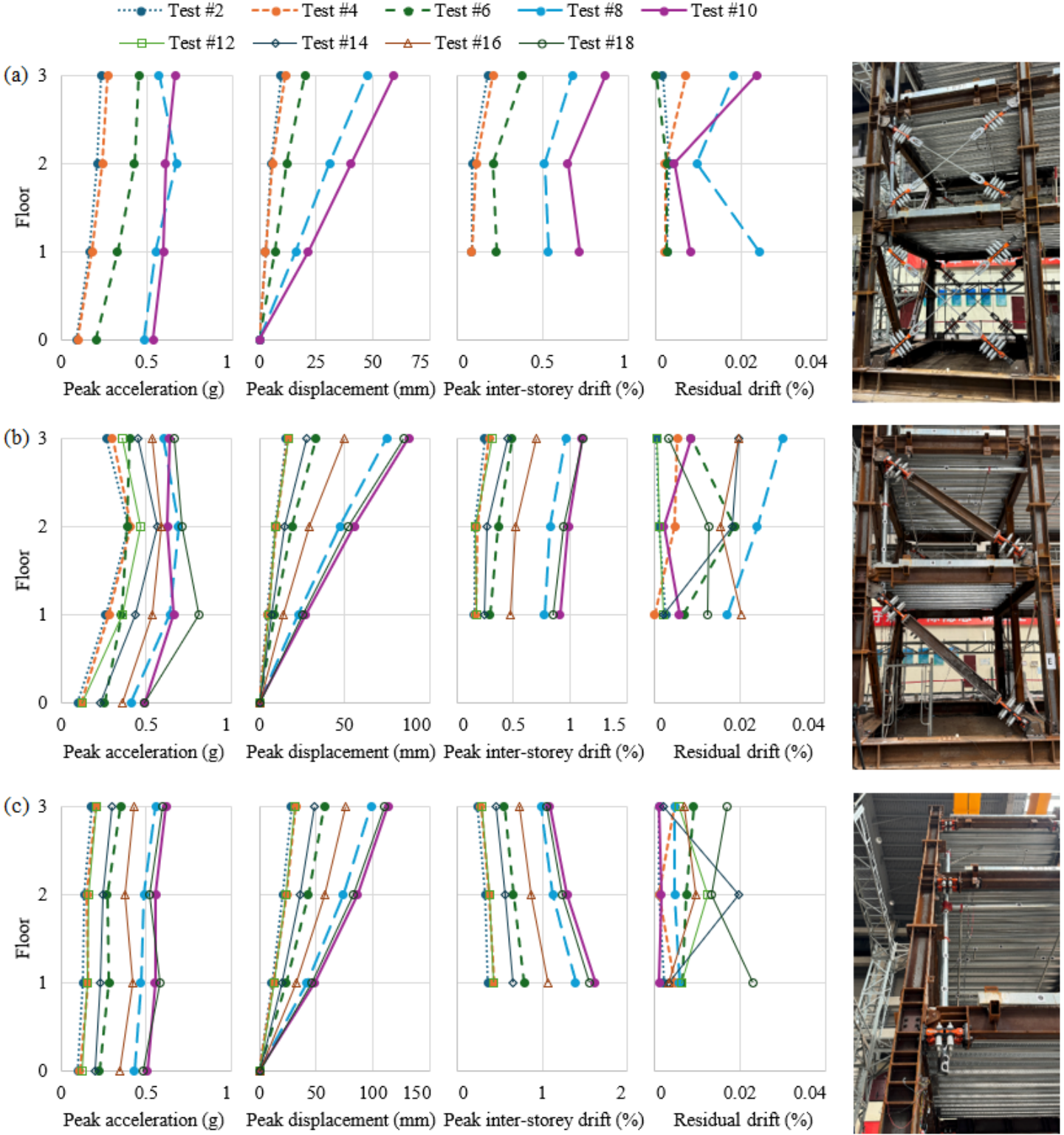

**Fig. 23.** Global responses of (a) TOB (b) TCB and (c) MRF systems. Note different scales used on the x-axes.

The deformation profiles (envelopes) were approximately linear for the brace concepts, thus achieving the intended design goal. For the moment-resisting frame, a curved profile was observed with the largest inter-storey drift concentrated in the lowest floor and progressively smaller drifts towards the top floor. The peak roof displacement was 59 mm (TOB), 88 mm (TCB) and 113 mm (MRF). This corresponds to 0.66%, 0.98% and 1.26% of the total height. The residual displacements were negligible for all the concepts tested. The largest value recorded across all the tests was 0.03%, which is well below the out-of-plumb threshold of 0.2% that is typically permitted for new construction (FEMA 2000). Hence, the RSFJ systems are effectively self-centering.

The peak absolute accelerations were similar throughout the structure for the MRF concept with only minor amplification of the PGA towards the top floor. On the other hand, some amplification was observed for the brace concepts. At the greatest intensity of unidirectional shaking (Test #10) the floor accelerations reached 0.67 g (TOB), 0.66 g (TCB) and 0.62 g (MRF) after averaging across the sensors on each floor. During bidirectional Test #18 of the TCB system, torsional effects may have caused unbalanced loading on the RSFJs and led to a larger spike of 0.81 g (averaged). Nevertheless, the displacement responses of the TCB and MRF systems were similar in both unidirectional and bidirectional tests as they increase accordingly with PGA.

**Table 7.** Global responses of the 3-storey building in terms of floor accelerations and displacements in the longitudinal ($x$) direction.

| Sys. | Test # | $PGA$ (g) | $a_{1m}$ (g) | $a_{2m}$ (g) | $a_{3m}$ (g) | $u_{1m}$ (mm) | $u_{2m}$ (mm) | $u_{3m}$ (mm) | $\delta_{1m}$ (%) | $\delta_{2m}$ (%) | $\delta_{3m}$ (%) | $\delta_{1res}$ (%) | $\delta_{2res}$ (%) | $\delta_{3res}$ (%) |
|------|------|------|------|------|------|------|------|------|------|------|------|------|------|------|
| TOB | 2 | 0.09 | 0.17 | 0.21 | 0.24 | 3 | 5 | 9 | 0.09 | 0.09 | 0.18 | 0.00 | 0.00 | 0.00 |
| | 4 | 0.10 | 0.18 | 0.24 | 0.27 | 3 | 6 | 12 | 0.09 | 0.11 | 0.21 | 0.00 | 0.00 | 0.01 |
| | 6 | 0.20 | 0.33 | 0.43 | 0.46 | 7 | 12 | 20 | 0.23 | 0.21 | 0.38 | 0.00 | 0.00 | 0.00 |
| | 8 | 0.49 | 0.56 | 0.68 | 0.57 | 16 | 31 | 47 | 0.53 | 0.51 | 0.67 | 0.02 | 0.01 | 0.02 |
| | 10 | 0.54 | 0.60 | 0.61 | 0.67 | 21 | 40 | 59 | 0.72 | 0.65 | 0.87 | 0.01 | 0.00 | 0.02 |
| TCB | 2 | 0.10 | 0.26 | 0.39 | 0.27 | 5 | 9 | 16 | 0.16 | 0.16 | 0.25 | 0.00 | 0.00 | 0.00 |
| | 4 | 0.12 | 0.28 | 0.41 | 0.30 | 5 | 10 | 17 | 0.17 | 0.18 | 0.29 | 0.00 | 0.00 | 0.01 |
| | 6 | 0.26 | 0.35 | 0.39 | 0.40 | 9 | 19 | 33 | 0.29 | 0.37 | 0.48 | 0.01 | 0.02 | 0.01 |
| | 8 | 0.41 | 0.64 | 0.69 | 0.60 | 23 | 47 | 75 | 0.77 | 0.82 | 0.96 | 0.02 | 0.02 | 0.03 |
| | 10 | 0.49 | 0.66 | 0.63 | 0.63 | 27 | 56 | 88 | 0.91 | 0.98 | 1.11 | 0.01 | 0.00 | 0.01 |
| | 12 | 0.13 | 0.36 | 0.47 | 0.36 | 5 | 10 | 17 | 0.18 | 0.17 | 0.32 | 0.00 | 0.00 | 0.00 |
| | 14 | 0.23 | 0.44 | 0.57 | 0.45 | 7 | 15 | 27 | 0.24 | 0.27 | 0.45 | 0.00 | 0.02 | 0.02 |
| | 16 | 0.36 | 0.54 | 0.59 | 0.54 | 14 | 29 | 50 | 0.47 | 0.52 | 0.70 | 0.02 | 0.02 | 0.02 |
| | 18 | 0.49 | 0.81 | 0.71 | 0.66 | 26 | 52 | 85 | 0.85 | 0.94 | 1.11 | 0.01 | 0.01 | 0.01 |
| MRF | 2 | 0.10 | 0.13 | 0.14 | 0.18 | 11 | 21 | 28 | 0.36 | 0.33 | 0.24 | 0.00 | 0.00 | 0.00 |
| | 4 | 0.11 | 0.16 | 0.15 | 0.21 | 13 | 24 | 32 | 0.43 | 0.38 | 0.28 | 0.00 | 0.00 | 0.00 |
| | 6 | 0.23 | 0.28 | 0.27 | 0.35 | 24 | 43 | 57 | 0.79 | 0.66 | 0.56 | 0.01 | 0.01 | 0.01 |
| | 8 | 0.43 | 0.47 | 0.49 | 0.55 | 42 | 74 | 99 | 1.39 | 1.13 | 0.99 | 0.01 | 0.00 | 0.00 |
| | 10 | 0.51 | 0.55 | 0.55 | 0.62 | 49 | 87 | 113 | 1.62 | 1.29 | 1.09 | 0.00 | 0.00 | 0.00 |
| | 12 | 0.12 | 0.15 | 0.16 | 0.21 | 13 | 24 | 32 | 0.43 | 0.38 | 0.29 | 0.01 | 0.01 | 0.01 |
| | 14 | 0.20 | 0.23 | 0.25 | 0.30 | 20 | 36 | 49 | 0.66 | 0.57 | 0.47 | 0.00 | 0.02 | 0.00 |
| | 16 | 0.35 | 0.42 | 0.38 | 0.43 | 32 | 57 | 76 | 1.07 | 0.86 | 0.74 | 0.00 | 0.01 | 0.01 |
| | 18 | 0.48 | 0.58 | 0.52 | 0.60 | 47 | 83 | 110 | 1.56 | 1.24 | 1.06 | 0.02 | 0.01 | 0.02 |

Note: Symbols $a, u, \delta$ denote the absolute acceleration, floor displacement and inter-storey drift. Subscripts $m, res$ denote the maximum and residual values, averaged across the sensors. Subscripts 1,2,3 denote the first, second and third floors.

*Displacement Histories*

The displacement histories, inter-storey drifts and the corresponding RSFJ deformation are shown in Figs. 24, 25 and 26 for the TOB, TCB and MRF concepts respectively. These are shown for Test #10 because the actual acceleration spectra provide a better match with the design spectrum and allows a fair comparison. In these figures, the RSFJ deformations shown for the brace concepts are the sum of the two joints along a brace and then averaged across all diagonals in the same storey and orientation. In the case of the MRF, the RSFJ deformations are averaged across all joints belonging to the same storey and transverse grid line (1 or 2).

All 3 systems consistently show Mode 1 deformation patterns during the largest vibration cycles and during the decay of vibrations. In the TOB and TCB tests, the deformations of the RSFJ appear to follow the inter-storey drift response. This indicates that the joints were performing as expected to resist and damp the shaking. The RSFJ deformation in the MRF tests was much lower than expected at a maximum of 1.9 mm. Additional flexibility in the system delayed the activation of the RSFJs and resulted in almost elastic behavior.

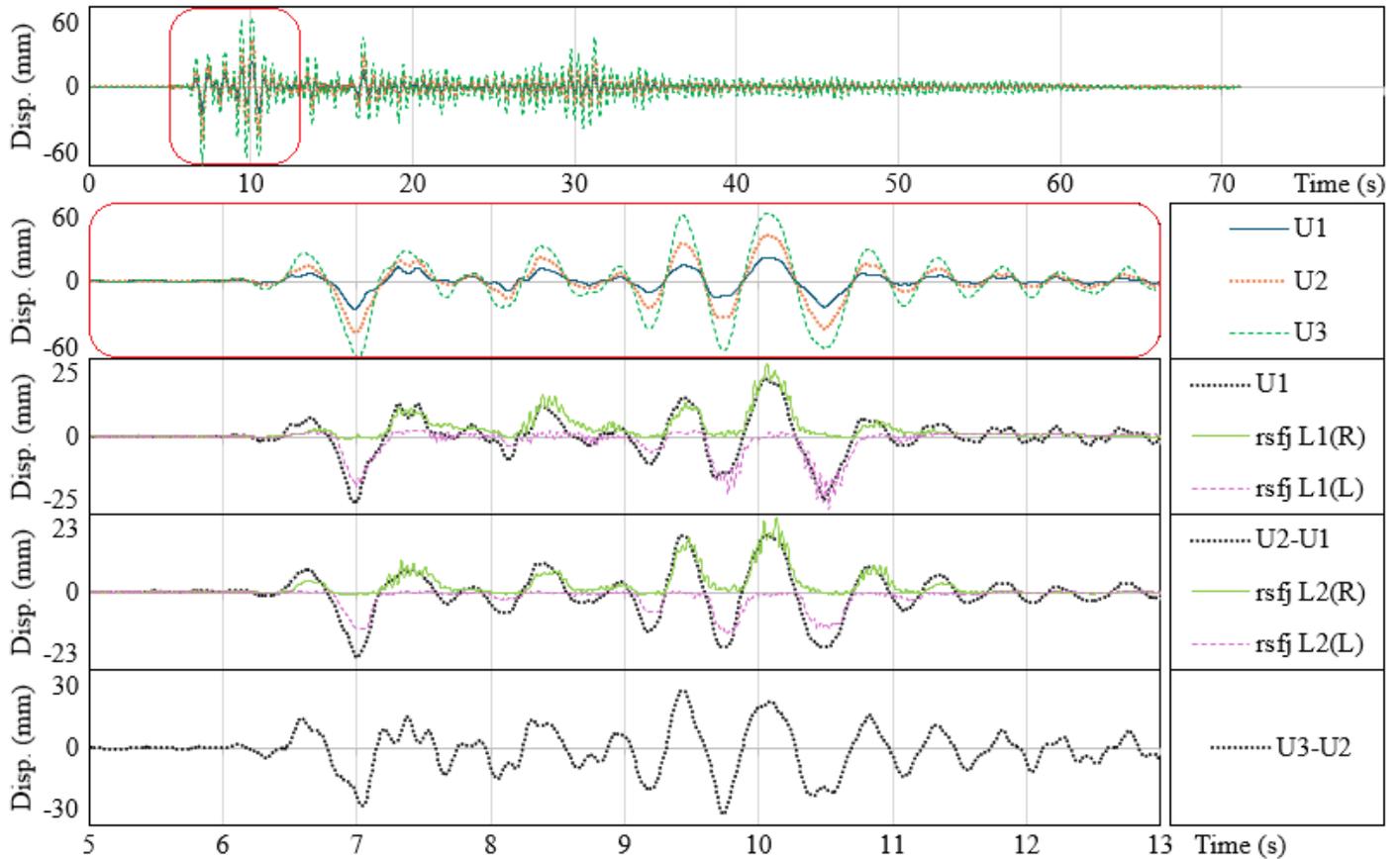

**Fig. 24.** TOB Test #10 displacement history.

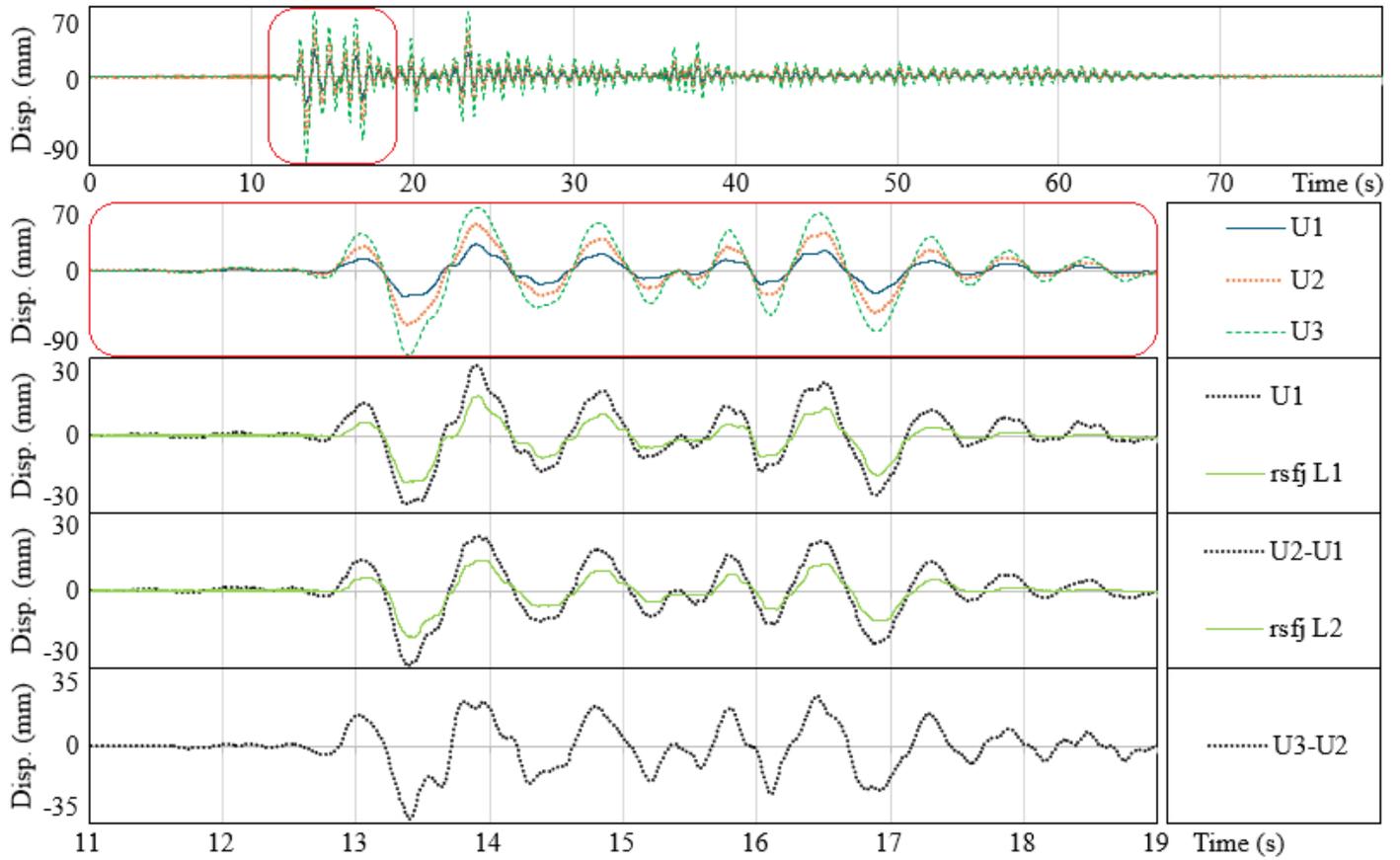

**Fig. 25.** TCB Test #10 displacement history.

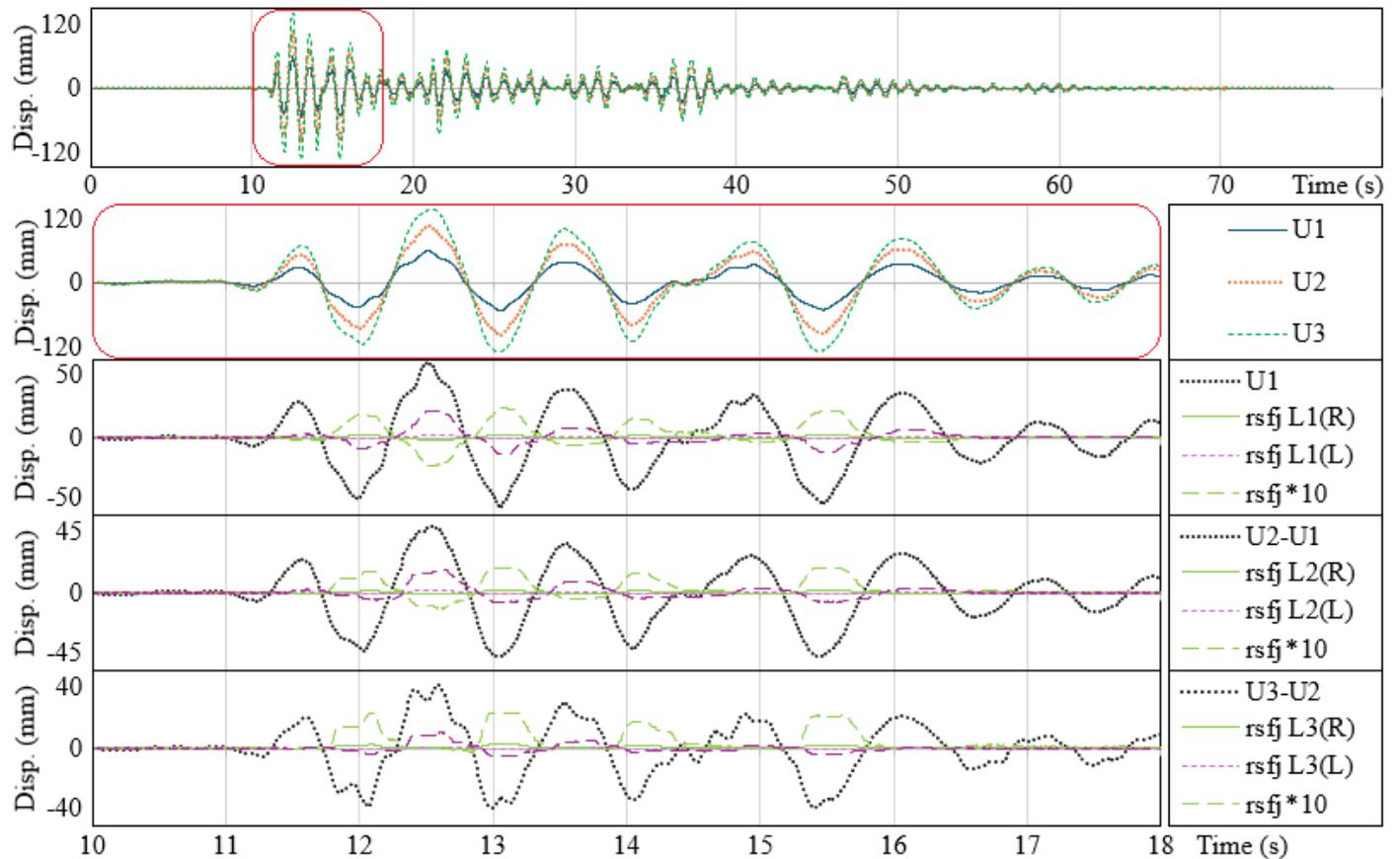

**Fig. 26.** MRF Test #10 displacement history.

In the TOB concept, the RSFJ deformations oscillated in a doubly periodic manner, exhibiting an additional higher frequency vibration with a period of 0.04 sec (25 Hz). This vibration is absent in the TCB concept which exhibited smoother RSFJ deformations. A possible cause is that the slender tension-only braces behave like a vibrating string. Notably, sideways motion occurs perpendicular to the brace axis when transitioning from a sagging state to a taut state, and the momentum contributes to a persisting sideways vibration – vibrations that are not damped when the joint is closed. The sideways deflection leads to corresponding axial fluctuations.

The frequency of the vibration observed in the tension-only brace (25 Hz) agrees with a rough estimate of the first harmonic frequency $f_1$ of 23 Hz calculated via Mersenne's law in Eq. 4. The calculation assumes a diagonal length of $L$ = 3.84 m from pin-to-pin, brace tension $F$ = 120 kN from Figs. 11 and 24, and a linear density $\rho$ = 3.9 kg/m with a 25 mm rod. In previous tests (Bagheri et al. 2020), the vibration was not significant because die springs were used to maintain constant tension (tautness) and minimize the sagging. As noted, the die spring resulted in a smoother transition by minimizing the impact loading when the joint engages the rod. Nevertheless, the vibrations did not translate into greater vibrations on a structure scale and thus did not have any appreciable impact on the performance of the system.

$$f_1 = \frac{1}{2L}\sqrt{\frac{F}{\rho}} \tag{4}$$

In general, the longitudinal displacements in bidirectional tests scale similarly with the PGA as in unidirectional tests. This is more so for the MRF system given its more elastic behavior. Figs. 27(b) and 26 show that the average longitudinal displacements were very similar between unidirectional and bidirectional tests. Minor differences between longitudinal grids A and E occurred because of a small torsion caused by the transverse system (as explained subsequently). However, the total rotation was approximately 3 times smaller compared to the TCB system that had more ductile joints. Hence, longitudinal displacements at grids A and E were relatively symmetric and synchronized for the MRF system. The supplemental materials show that the peak displacements match very closely.

In the TCB system, bidirectional Test #18 showed visible torsion due to unbalanced resistance in the transverse direction. Specifically, the $y$-displacements were larger on grid 1-1 as the SFC deformed more than those on grid 3-3. Grid 1-1 was especially flexible on Storeys 2 and 3, such that torsion initiated at these floors just after 12.8 sec. The rotation history in Fig. 27(a) and displacement trace in Fig. 28 show that the twisting oscillated between clockwise and counterclockwise directions in between the intervals of (a) 12.95–13.15 sec, (b) 13.22–13.45 sec, (c) 13.45–13.60 sec, and (d) 13.60–13.67 sec, depending on the relative $y$-displacement and velocity to the ground. These torsional vibrations are also reflected in the SFC and RSFJ deformations and their differences between the grids. In contrast to the uniform $x$-displacements of unidirectional Test #10, the existing rotation in Test #18 meant that $x$-displacements at the corners A1/E3 were slightly more positive/negative (for counterclockwise rotation) or negative/positive (for clockwise rotation). On average however, the peak displacements only differ slightly between Tests #10 and #18 (Table 7). The torsion also had a noticeable impact on TCB floor accelerations as described in the next section.

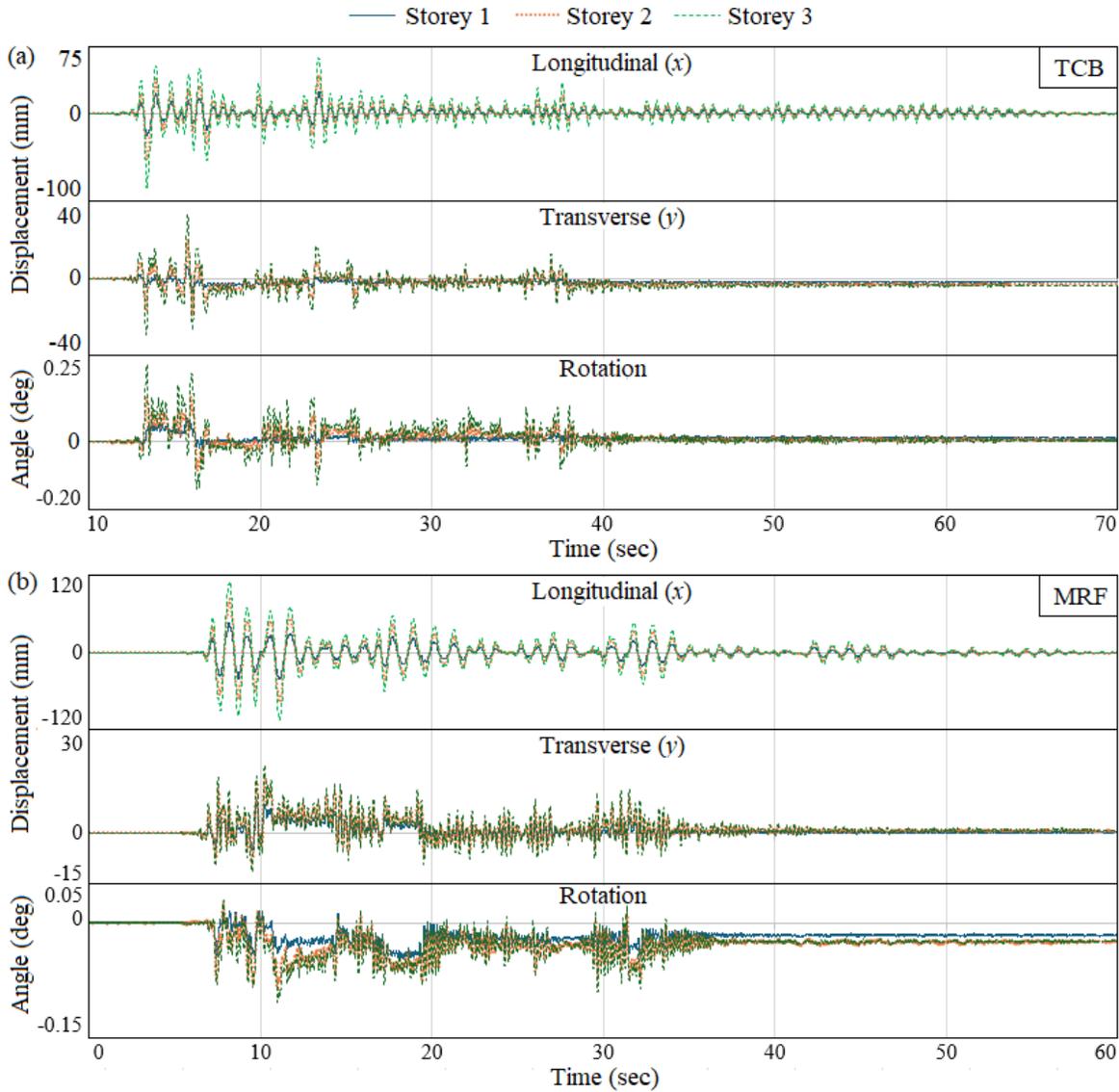

**Fig. 27.** Relative displacements and floor rotations during bidirectional Test #18 for (a) TCB system, and (b) MRF system.

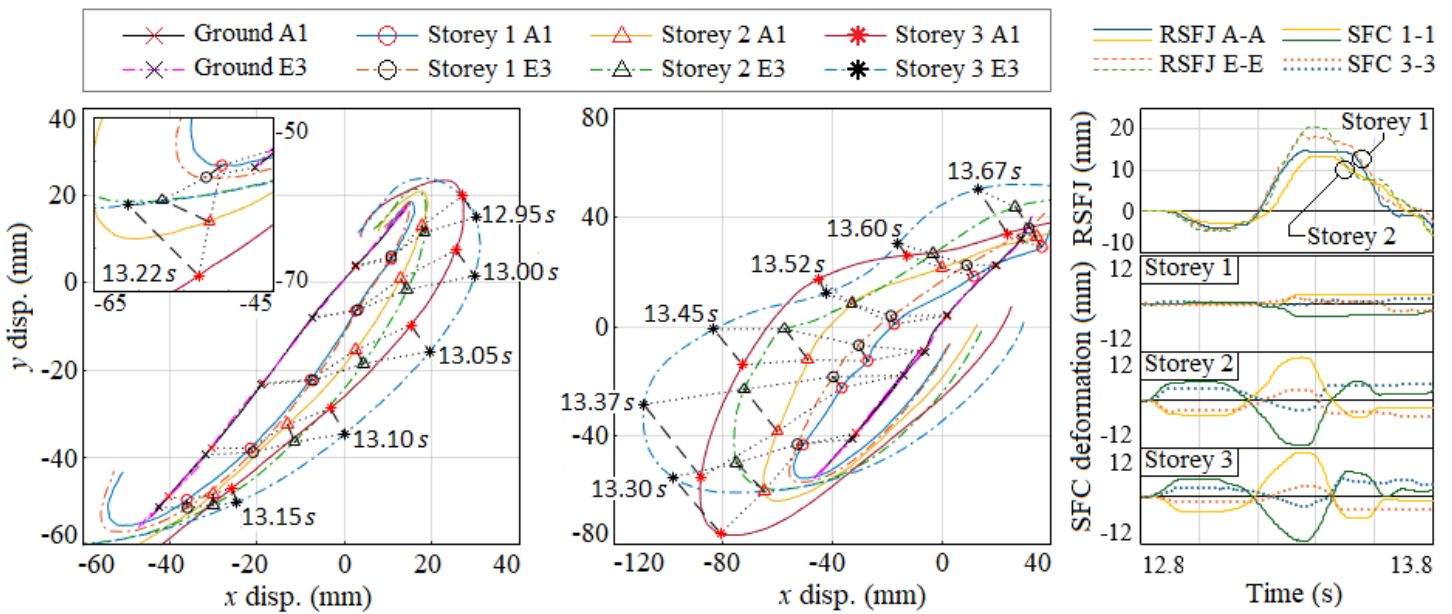

**Fig. 28.** Trace of floor (absolute) displacements for TCB Test #18 between 12.8–13.8 sec.

*Floor Accelerations*

Systems like RSFJ aim to prevent permanent damage to the structural skeleton and the risk of significant downtime. Despite that, time and financial losses may arise from shaking-induced damage to valuable non-structural contents of a building. These components are known to be acceleration sensitive like toppled objects. Hence, this section describes the floor accelerations in more detail.

At the top floor, the acceleration histories appear to mirror the inter-storey drift responses very closely (Fig. 29). Fluctuations in acceleration occur alongside slope (velocity) changes in the inter-storey drift responses. Furthermore, the timings of the peak accelerations coincide with the peak inter-storey drifts, i.e., during reversal of motion at the peak amplitudes. These indicate harmonic behavior at the top floor as the drift and acceleration responses are in phase with each other. This result is somewhat expected given the fact that all 3 concepts are essentially elastic at the top floor, due to the lack of braces here and the low slip of the MRF joints.

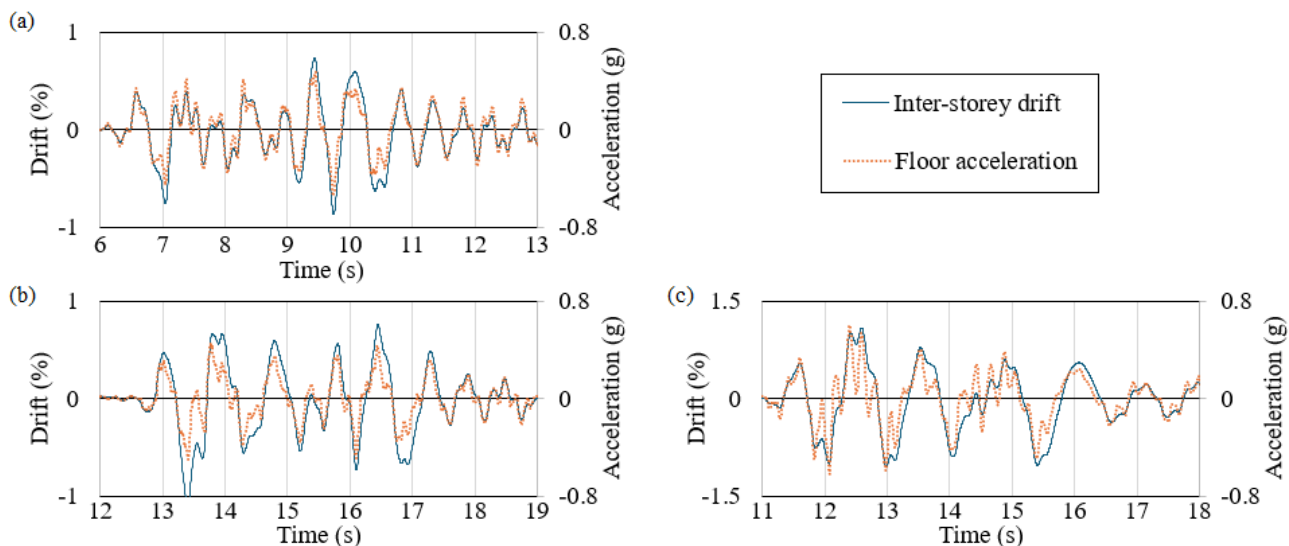

**Fig. 29.** Top floor drift and acceleration responses from Test #10 of the (a) TOB (b) TCB and (c) MRF systems.

In TOB Test #10, a higher frequency Mode 2 occurred simultaneously with a broader Mode 1 pattern between 7.2 sec to 7.6 sec in Fig. 30(c). Despite the appearance of a Mode 1 profile, the first two floors vibrated locally by approaching the ground and top floors alternatingly, with L1 and L2 being out of phase with L0 and L3. The resonating oscillations led to acceleration spikes at the $1^{st}$ floor (7.45 sec, 0.642 g) and at the $2^{nd}$ floor (7.48 sec, 0.517 g) even though the preceding ground acceleration was only 0.32 g. The low drifts are additional indicators of the presence of a higher mode.

Spikes were also registered at around 31 sec in Fig. 30(d) during the weaker cluster of shaking, at approximately 3 times amplification of the preceding ground acceleration. The cause of the $1^{st}$ floor spike at 30.84 sec (0.434 g) was due to the stiffness transition, which occurred at the instance of zero crossing during Mode 1 vibration. This is explained in the next paragraph. For the $2^{nd}$ floor spike at 31.58 sec (0.403 g), it occurred at peak displacement amplitude when the ground and top floors reversed direction sharply, followed by the first and second floors reversing in sync. This pattern of vibration with the first two floors deflecting in sync resembles the multiple-mode vibration noted at the earlier spikes between 7.2–7.6 sec.

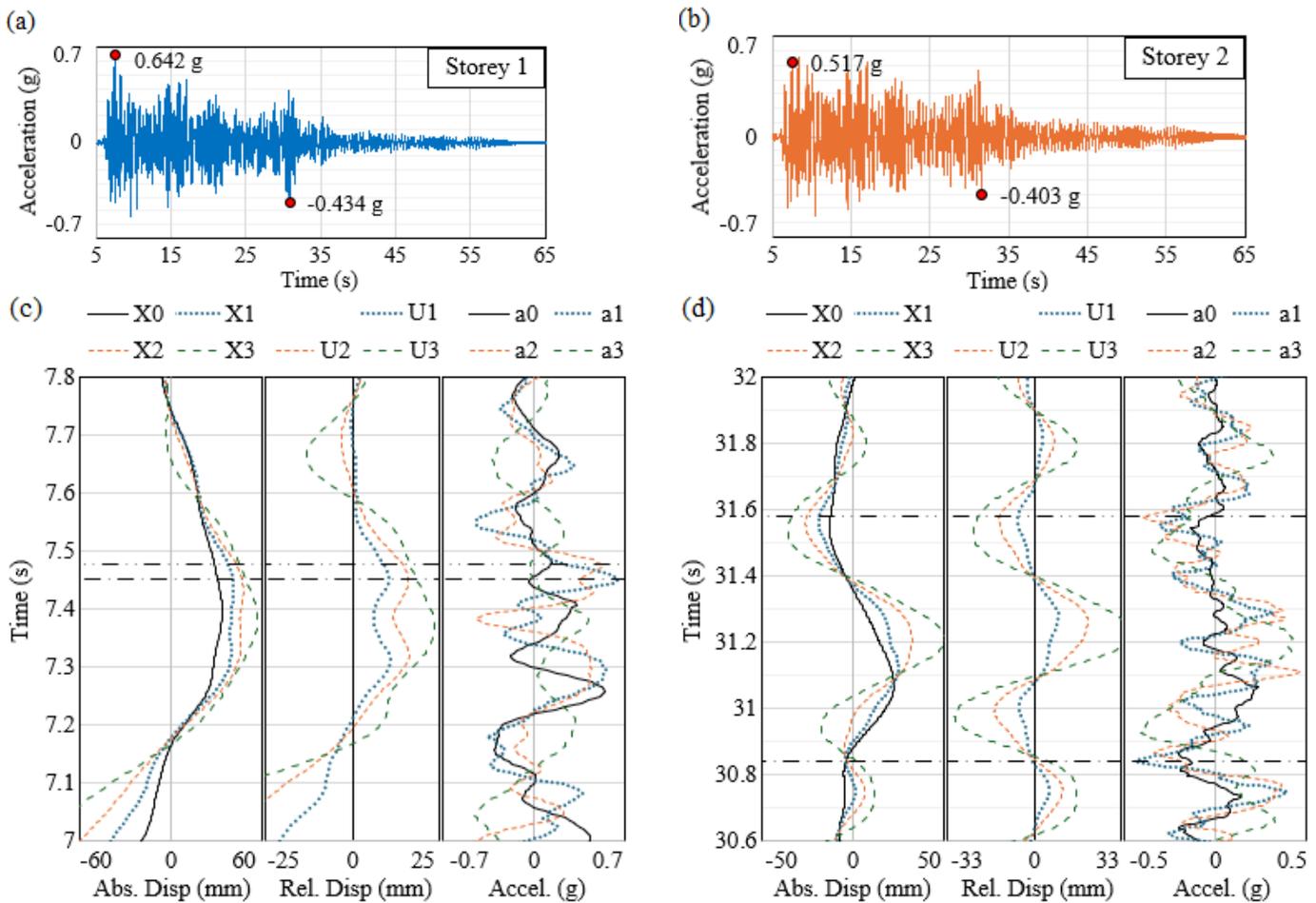

**Fig. 30.** TOB Test #10 responses in (a) first and (b) second storeys. (c) First spike and (d) second spike for both floors.

Prior numerical modelling has indicated the possibility of acceleration spikes associated with the sudden transition of stiffness in self-centering mechanisms. When the system re-centers at high speed (aided by restoring forces) it may suddenly 'lock up' when it crosses over the zero position and encounters a high initial stiffness that resists motion in the opposite direction. Thus, the acceleration spike results from the sudden change in velocity of the floor. Erochko et al. (2013) found that this phenomenon did not manifest as severely in experimental tests at about one half of the modelled value. A possible reason is a more gradual transition due to the flexibility of the many components in a system, resulting in a more gradual acceleration or deceleration. Nevertheless, the largest floor acceleration was recorded at the instance of zero crossing, proving the existence of acceleration spikes caused by self-centering joints.

In TCB Test #10, an acceleration spike associated with the stiffness transition can also be seen during the first cluster of intense shaking at approximately 13 seconds in Fig. 31(c). Although the structure was vibrating in the first mode, the spikes occurred near zero inter-storey drift. This contrasts with the (top floor) spikes at peak drift during a reversal of direction. In this case, the table decelerated sharply, and spikes were recorded at the first floor (13.64 s, 0.649 g) and almost immediately after at the second floor (13.68 s, 0.596 g). The timing of the spikes coincided with the closure of the joints near zero inter-storey drift. This suggests that the stiffness regained in the intermediate floors made them susceptible to the sudden deceleration of the table.

Significant amplification of acceleration was also detected in the first two floors (L1 and L2) during the final and weakest cluster of shaking around 35 seconds, despite the ground reaching only one-third of the PGA. The cause can be traced to a higher mode (Mode 2) response vibrating at an approximate period of 0.2 sec, which agrees with the period in Table 5. Fig. 31(d) shows the transition from Mode 1 to Mode 2 behavior that built up to the spike in the first storey (35.61 sec, 0.631 g). The sudden deceleration of the first floor was caused by the ground deceleration and the momentum of the top floor, both of which induced a sharp pull on L1 towards the right. In the lead up to the second storey spike (35.83 sec, 0.568 g), L2 was at the maximum displacement amplitude and both the adjacent floors L1 and L3 were pulling it in the opposite direction – L3 pulling via restoring forces, and L1 aided by the ground acceleration towards the right. Combined, the two forces pulled L2 to the positive direction with great acceleration akin to a 'slingshot' effect. The pull of the ground was large enough to return the structure to Mode 1 vibration pattern.

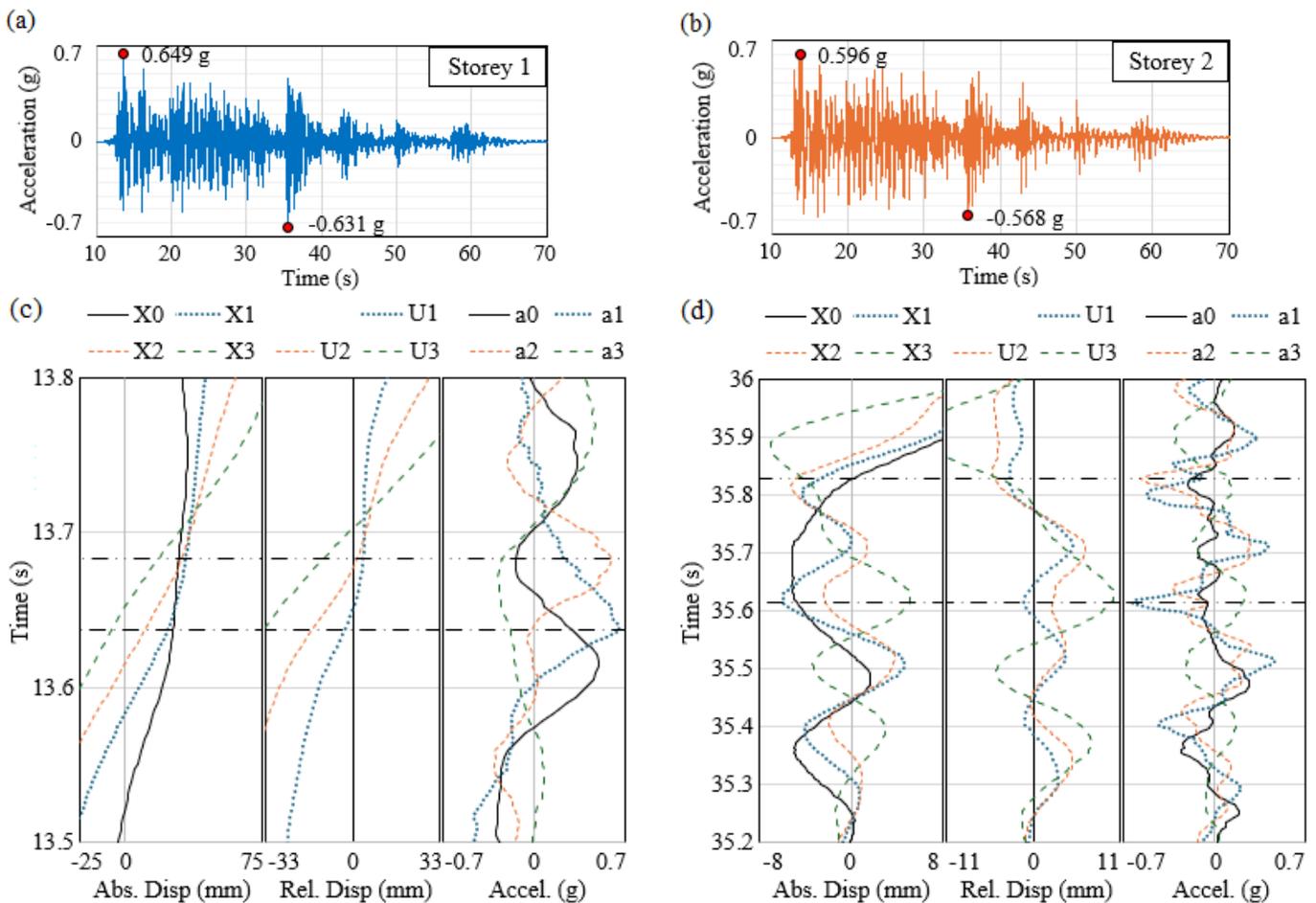

**Fig. 31.** TCB Test #10 responses in (a) first and (b) second storeys. (c) First spike and (d) second spike for both floors.

In MRF Test #10, the structure behaved in a more elastic fashion as there was minimal slip in the joints. Hence, the first and largest set of spikes at the 1st floor (12.49 sec, 0.547 g) and the 2nd floor (12.52 sec, 0.540 g) occurred during a reversal of direction at the peak deflection of a Mode 1 profile (Fig. 32). The second set of spikes at the 1st floor (33.93 sec, 0.383 g) and the 2nd floor (34.15 sec, 0.255 g) were also caused by a Mode 2 vibration even though they were relatively smaller for the MRF. It is apparent that the last cluster of shaking has a propensity to excite a higher mode for all 3 systems.

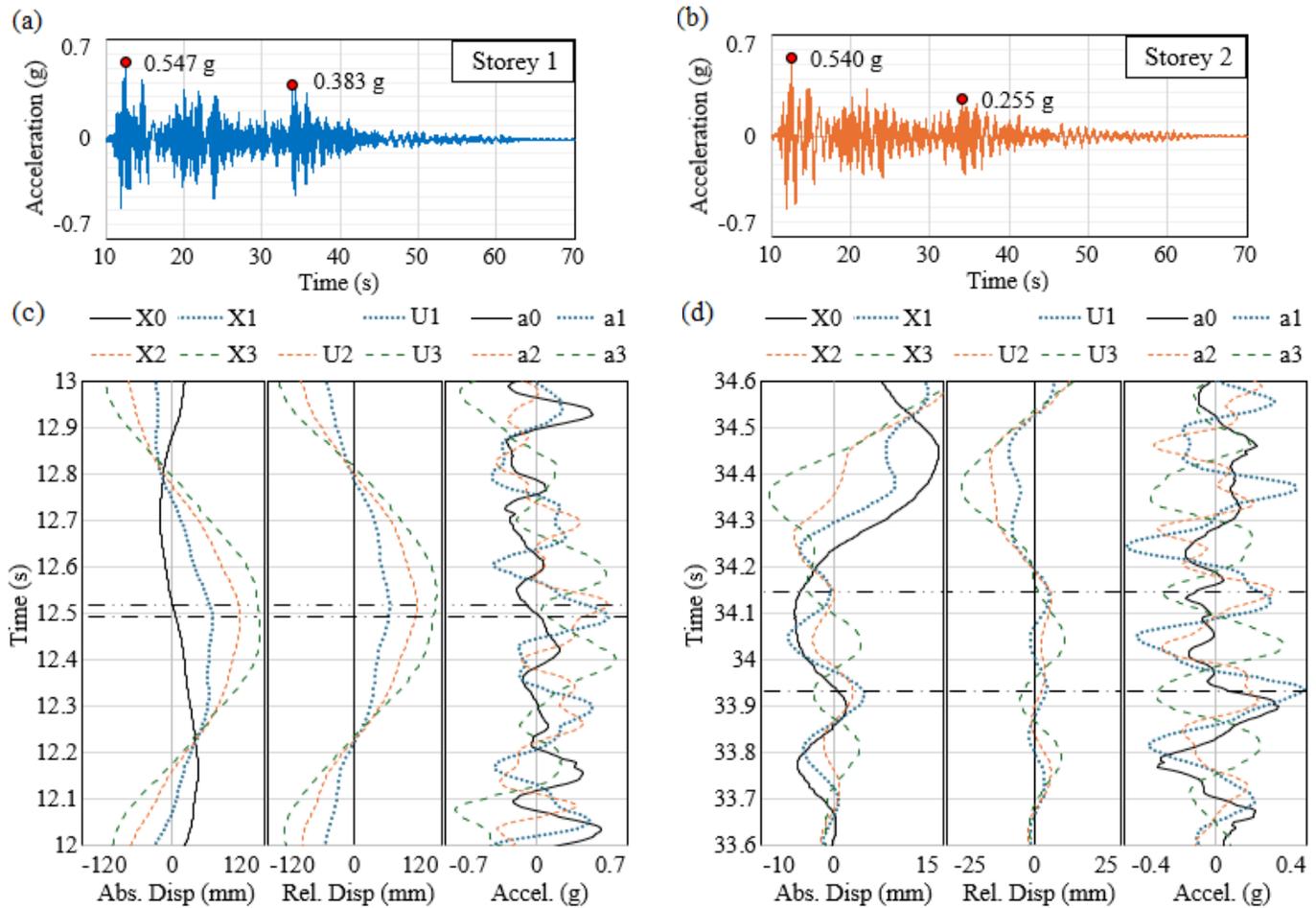

**Fig. 32.** MRF Test #10 responses in (a) first and (b) second storeys. (c) First spike and (d) second spike for both floors.

In the bidirectional TCB test, the previous acceleration spike of 0.65 g from Test #10 has increased substantially to 0.95 g in Test #18 (Fig. 33). This is the largest floor acceleration recorded at an amplification of 2 times the PGA (0.49 g). However, the acceleration spike was not uniform across the diaphragm as it ranged from 0.73 g at column E1 to 0.95 g at column A3. A closer examination of the spike at 13.66 sec shows that it occurred during a stiffness transition (as in Test #10), but it was exacerbated by torsional effects causing greater (uneven) loading on grid line A. The displacement trace in Fig. 28 earlier shows that grid line A was further to the right compared to grid line E around 13.6 sec and allow RSFJs on grid A to begin the stiffness transition earlier.

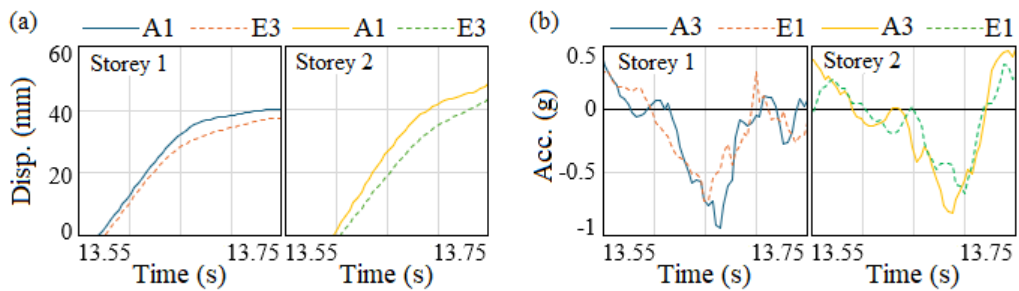

**Fig. 33.** Torsion effects on floor (a) displacements and (b) accelerations during TCB Test #18.

The unequal loading is observed in the RSFJ deformation history in Fig. 28, which also shows smaller deformations on grid A. As the floor diaphragm possesses some rigidity, both accelerometers started decelerating at the same time due to RSFJs on grid A (Fig. 33). However, the deceleration on grid A continued to increase with continued loading of the RSFJs on this grid. On the other hand, grid E was trailing behind grid A, where both Storeys 1 and 2 have already crossed the zero point in Fig. 28. Hence, as shown in Fig. 33(a), the floors decelerated sharply at grid A in comparison with grid E which had room for a more gradual deceleration.

**Conclusions**

This study presents shake table tests of 3 different seismic resisting systems that utilize the RSFJ, as part of (1) tension-only braces (TOB), (2) tension-compression braces (TCB), and (3) moment-resisting frame (MRF) joints. All 3 systems were subjected to increasing levels of shaking by scaling the 1940 El Centro ground motion up to a PGA of 0.5 g. In all cases, the systems demonstrated self-centering behavior effectively. The largest residual displacement recorded across all tests was 0.03%, which is well below the out-of-plumb threshold of 0.20% typically permitted for new builds.

All systems satisfied the design requirements in terms of lateral displacements. At the largest intensity of shaking, the peak displacements recorded at the top floor were 0.66%, 0.98% and 1.26% of the total height for the TOB, TCB and MRF systems respectively. The peak inter-storey drift occurred at the top floor for the braced systems (TOB: 0.87%, TCB: 1.11%) and it was concentrated at the lowest floor for the moment-resisting frame (MRF: 1.62%). While the RSFJs in the TOB and TCB damped the vibrations effectively, the RSFJs in the MRF system did not slip as much as intended due to the flexibility of the frame. This resulted in deflection envelopes with approximately linear profiles for the TOB and TCB systems and a curved profile for the MRF system.

During the strongest shaking, the peak accelerations reached 0.67 g (TOB), 0.66 g (TCB) and 0.62 g (MRF). Top floor accelerations tend to peak in tandem with the inter-storey drift (i.e., during reversal of direction at the maximum drift), resembling harmonic vibration because of elastic behavior at the top floor. At the intermediate floors, acceleration spikes occurred due to a sudden transition of stiffness in the braced systems. After re-centering at high speed, a sudden change in slip resistance at the zero/plumb position caused sharp deceleration. This contrasts with top floor spikes at the peak amplitude, despite both vibrating in Mode 1.

A higher mode (Mode 2) vibration pattern can occur either alone or simultaneously with a broader Mode 1 pattern, as observed in the TOB system. By itself, the Mode 2 vibrations amplified the floor accelerations at 2–3 times the ground acceleration in the associated cycle. However, these occurred during the latter and weaker cycles of ground motion. In absolute terms, these peaks were smaller and secondary to the major spikes caused by the first and strongest cluster of shaking.

Finally, torsion effects can compound acceleration spikes, resulting in the largest accelerometer reading of 0.95 g (2x PGA). During the strongest bidirectional shaking, the TCB system experienced 3 times greater torsion compared to the more elastic MRF system. The torsion was caused by unbalanced resistance of connections in the transverse direction, leading to uneven loading of the longitudinal grid lines that contain the RSFJs. In terms of peak displacements however, the longitudinal translation of the floors only deviated up to 7% from the unidirectional test. Hence, the bidirectional tests show that the RSFJ systems can perform effectively even during simultaneous out-of-plane shaking.

This paper shows that RSFJs can be used in various structural systems effectively against seismic hazards, as they can (1) limit the inter-storey drifts to values below code-permitted levels, thus reducing damage to structural and non-structural components, and (2) prevent residual drifts and avoid costly realignment repairs, thus enabling a faster recovery after earthquakes and improving the resilience of communities. Further avenues for research include investigating floor accelerations to minimize the risk of damage to building contents and occupants. The results also indicate opportunities to improve the MRF concept and exploit the full capabilities of the RSFJ damper.

## Data Availability Statement

Some or all data, models, or code that support the findings of this study are available from the corresponding author upon reasonable request.

## Acknowledgements

This work was made possible through the contributions of various organizations from New Zealand (NZ) and China, including the International Joint Research Laboratory of Earthquake Engineering (ILEE) at Tongji University, Building Research Association of NZ (BRANZ), Heavy Engineering Research Association (HERA), QuakeCoRE NZ Centre for Earthquake Resilience, the Building Innovation Partnership, NZ Earthquake Commission (EQC), NZ Tertiary Education Commission, NZ Ministry of Business & Innovation and Employment (MBIE), University of Auckland (UoA), Auckland University of Technology (AUT) and University of Canterbury (UC). Material donations are acknowledged from ComFlor, Hilti Corporation, and Lanyon & LeCompte Construction Ltd.